\documentclass[aps,prc,twocolumn,groupedaddress,a4paper]{revtex4-1}
\usepackage{amsmath}
\usepackage{amssymb}
\usepackage{graphicx}
\usepackage{textcomp}
\usepackage{braket}
\usepackage{tabularx}
\usepackage{booktabs, hhline}
\usepackage[dvipsnames]{xcolor}

\newcommand {\la} {\langle}\newcommand {\ra} {\rangle}
\newcommand {\beq} {\begin{eqnarray}}
\newcommand {\eeqn} [1] {\label{#1} \end{eqnarray}}
\newcommand {\eol} {\nonumber \\}
\newcommand {\ve} [1] {\mbox{\boldmath $#1$}}

\begin{document}

\title{
Three-body optical potentials in  $(d,p)$ reactions and their influence on indirect study of stellar nucleosynthesis}

\author{M. J. Dinmore}
\affiliation{Department of Physics, Faculty of Engineering and Physical
Sciences, University of Surrey, Guildford, Surrey, GU2 7XH, United Kingdom}
\author{N. K. Timofeyuk}
\affiliation{Department of Physics, Faculty of Engineering and Physical
Sciences, University of Surrey, Guildford, Surrey, GU2 7XH, United Kingdom}
\author{J.S. Al-Khalili}
\affiliation{Department of Physics, Faculty of Engineering and Physical
Sciences, University of Surrey, Guildford, Surrey, GU2 7XH, United Kingdom}

\date{\today}

\begin{abstract}
Model uncertainties, arising due to suppression of target excitations in the description of deuteron scattering and resulting in a modification of the two-body interactions in a three-body system, are investigated for several $(d,p)$ reactions serving as indirect tools for studying the  astrophysical $(p,\gamma)$ reactions relevant to $rp$-process. The three-body nature of deuteron-target potential is treated within adiabatic distorted wave approximation (ADWA) which relies on dominant contribution from the components of the three-body deuteron-target wave function with small $n$-$p$ separations. This results in a simple prescription for treating the explicit energy-dependence of two-body optical potentials in a three-body system requiring  nucleon optical potentials to be evaluated at a shifted energy with respect to the standard value of half the deuteron incident energy. In addition, the ADWA allows for leading-order  multiple scattering effects to be estimated, which leads to a simple renormalization of the adiabatic potential's imaginary part by the factor of two. These effects are assessed using both nonlocal and local optical potential systematics for $^{26}$Al, $^{30}$P, $^{34}$Cl and $^{56}$Ni targets at deuteron incident energy of 12 MeV typical for experiments with radioactive beams in inverse kinematics. The model uncertainties induced by the three-body nature of deuteron-target scattering are found to be within 40$\%$  both in the main peak of angular distributions and in total $(d,p)$ cross sections. At higher deuteron energies, around 60 MeV, model uncertainties can reach 100\% in the total cross sections. A few examples of application to astrophysically interesting proton resonances  in $^{27}$Si and $^{57}$Cu obtained using $(d,p)$ reactions and mirror symmetry are given.
\end{abstract}
\maketitle

\section{Introduction.}



%

In order to fully describe the relative abundances of the elements we must be able to properly constrain the reaction rates of the stellar processes that drive them. 
One class of reactions important for understanding stellar nucleosynthesis is proton capture, $(p,\gamma)$.
Measuring $(p,\gamma)$ cross sections directly is a difficult task since they are very small. 
One technique for the investigation of exotic nuclei produced by $(p,\gamma)$ in stellar 
environments is to perform deuteron-induced one-nucleon transfer reactions $(d,n)$ and $(d,p)$ with radioactive beams, utilising inverse kinematics. 
The first reaction, $(d,n)$, contains the same crucial piece of information - the overlap function of the wave functions of nuclei in initial and final states - as the $(p,\gamma)$ reaction does.
Populating bound states in $(d,n)$ reactions provides  these overlap functions 
in a straightforward manner. However, in many cases, such as the $rp$ process, astrophysically important states are unbound, and their indirect access via $(d,n)$ reaction encounters problem both on the experimental side, where neutron detection could be problematic, and the theoretical side, where the transfer to continuum is not yet well understood. The corresponding mirror states that could be populated in $(d,p)$ reactions are bound and more easily accessible to study, both experimentally and theoretically. The information about widths $\Gamma_p$ of proton states populated in $(p,\gamma)$ reactions could be obtained from the relations between these widths and the Asymptotic Normalization Coefficients (ANCs) of the their mirror states populated in $(d,p)$ reactions \cite{Tim03a}. The ANCs determine the magnitude of the asymptotic decrease of the overlap function.

Extracting ANCs from $A(d,p)B$ reactions takes place through comparison of measured angular distributions with those calculated with the help of  reaction theory. 
Given that deuteron breakup effects are important, their effects are often accounted for in the Adiabatic Distorted Wave Approximation (ADWA) \cite{JT}. This theory is based on a $n+p+A$ three-body description of the deuteron-target motion and involves pairwise $n$-$p$, $p$-$A$, and $n$-$A$ interactions. The latter two are assumed to be given by nucleon optical potentials taken at half the deuteron incident energy $E_d$. In reality, the $A+n+p$ system is a complex many-body system, and it was shown that projecting its wave function onto a three-body channel results in a complicated three-body optical operator that includes multiple scattering to all orders \cite{Joh14}. Even the simplest, leading order, terms of this operator do not look like $n$-$A$ and $p$-$A$ optical potentials since they explicitly depend on the position of the third particle ($p$ or $n$) and on the three-body rather than two-body energy. It has been possible to estimate the contribution of the leading terms of this operator within ADWA \cite{Joh14}, which requires using nonlocal energy-dependent optical potentials taken at an energy shifted from the traditionally used value of $E_d/2$ by $n$-$p$ kinetic energy averaged over the short range of the $n$-$p$ force. Adding the first-order term of the three-body optical operator in a leading order results in a doubling the dynamical part of this optical potential \cite{Din19}.

The present paper aims to quantify uncertainties, arising due to induced three-body force, of model predictions of $(d,p)$ reactions used to indirectly probe $(p,\gamma)$ capture that occurs in $rp$ process. We assess these effects within the ADWA. In section II we summarise the ADWA three-body optical potential formalism of \cite{Joh14} and \cite{Din19}, arising due to neglecting the channels with target excitations.  We will apply this formalism to  $(d,p)$ reactions, listed in section III, that could serve as indirect tools for measuring proton capture reactions of astrophysical interest. In section IV we discuss numerical results obtained with different nucleon-nucleon (NN) models for a global nonlocal nucleon optical potential both with and without the induced leading-order multiple scattering three-body   force, and we also explore the possibility of using local optical potentials within this approach. In section V we study model uncertainties due to the three-body optical potential in total $(d,p)$ cross sections. In section VI we apply mirror symmetry to determine the proton width of some astrophysically relevant states in $^{27}$Si and $^{56}$Co using their relations   to the ANCs of their mirror states derived from ($d,p)$ reactions. Conclusions are formulated in section VII highlighting the necessity and directions for future research.

\section{Induced three-body optical potentials within the ADWA formalism.}

It has been shown in \cite{Din19} that projecting the many-body wave function of the $A+n+p$ system onto a three-body channel results in a three-body Hamiltonian 
\beq
H_3=T_3+V_{np}+V_{\text{opt}},
\eeqn{H3}
where $T_3$ is the three-body kinetic energy operator, $V_{np}$ is a short-range $n$-$p$ interaction and $V_{\text{opt}}$ is a complicated three-body optical potential. The latter is the target-ground-state expectation value $\la \phi_A| U| \phi_A\ra$ of the optical  operator $U$ that contains multiple scattering to all orders:
\beq
U &=& \underbrace{U_{nA}+U_{pA}}_\text{$U^{(0)}$}+
\underbrace{U_{nA}\dfrac{Q_A}{e}U_{pA}+U_{pA}\dfrac{Q_A}{e}U_{nA}}_\text{$U^{(1)}$}
\eol
&+& \underbrace{U_{nA}\dfrac{Q_A}{e}U_{pA}\dfrac{Q_A}{e}U_{nA}+U_{pA}\dfrac{Q_A}{e}U_{nA}\dfrac{Q_A}{e}U_{pA}}_\text{$U^{(2)}$}+\cdots
\eol
\eeqn{U}
where
\beq
U_{NA} = \left(1 -v_{NA} \frac{Q_A}{e} \right)^{-1} v_{NA}.
\eeqn{UNA}
Here, the operator $Q_A $  projects onto the model space defined by all excited states of $A$ and excluded from consideration, $e = E_{3}+i0-T_{3}-V_{np}-(H_{A}-E_{A})$ and $E_3$ is the three-body energy in the $A+n+p$ system, while $v_{NA}$ is the sum of interactions of nucleon $N$ (either $n$ or $p$) with nucleons of the target $A$. The ground state wave function $\phi_A$ is defined by the 
many-body Hamiltonian $H_A$, $H_A \phi_A = E_A \phi_A$, with $E_A$ being the intrinsic binding energy of the target.

The ADWA model starts from  expanding the eigenfunction $\Psi(\ve{R},\ve{r})$ of the three-body  Hamiltonian $H_3$ over the Weinberg basis functions $\phi_i(\ve{r})$,
\begin{equation}
[-\epsilon_{d}-T_{r}-\alpha_{i}V_{np}]\phi_i(\ve{r}) = 0, \qquad i=1,2,...,
\label{eigen}
\end{equation}
where $\ve{r}$ is the $n$-$p$ separation coordinate and $T_r$ is the kinetic energy operator associated with it. The $\phi_i$ satisfy the orthonormality relation
\beq
\braket{\phi_i|V_{np}|\phi_j}=-\delta_{i,j}.
\eeqn{orthonormality}
The ADWA retains only the first term of the Weinberg-state expansion of $\Psi(\ve{R},\ve{r})$ \cite{JT}. This results in the $\Psi(\ve{R},\ve{r}) \approx \chi_{dA}^{(+)}(\ve{R}) \phi_d(\ve{r})$ approximation with $\phi_d$ being the deuteron ground state wave function and the $d$-$A$ channel distorted wave function $\chi_{dA}^{(+)}$ satisfying the two-body Schr\"odinger equation
\beq
\left(T_R +V^{\rm ADWA}(R) 
-E_d\right) \chi_{dA}^{(+)}(\ve{R}) = 0, 
\eeqn{2bAD}
where $T_R$ is the deuteron kinetic energy operator associated with the relative coordinate $\ve{R}$ connecting the centres of mass of the deuteron and the target $A$. The adiabatic potential $V^{\rm ADWA}$ is the first diagonal
matrix element of the optical operator $U$, which due the Weinberg orthonormality condition is defined as 
$\la \phi_1| U | \phi_d\ra$, where 
\beq
\phi_1 = \frac{V_{np}\phi_d}{\la\phi_d|V_{np}|\phi_d\ra}.
\eeqn{phi1}
The matrix element of the first term, $U^{(0)}$,  of the operator $U$   contains  interactions of $n$ and $p$ with the target nucleons only, so that it can be associated with the $n$-$A$ and $p$-$A$ optical potentials. All other terms will contain interactions between $n$ and $p$ via excitation of excited states in $A$. The associated matrix elements 
will combine into a three-body potential. Including it into the model Hamiltonian will account for three-body effects induced by the target excitation. We call them induced three-body effects (I3B).

It has been shown in \cite{Din19} that in the ADWA, the simplest term, $U^{(0)} =U_{pA} +U_{nA}$,  of the operator $U$ 
is given by $U_{nA}^{\rm ADWA}+U_{pA}^{\rm ADWA}$, with 
\beq
U_{NA}^{\rm ADWA}&=&v_{NA}\eol&+&v_{NA}\frac{Q_A}{E_{\text{eff}}-T_N-H_A-Q_Av_{NA}Q_{A}}v_{NA}, \,\,\,\,\,\,\,\,\,\,\,\,
\eeqn{UNA}
where  
\beq
E_{\rm eff} &=& E_d/2 + \Delta E 
\\
\Delta E &=& \frac{1}{2}   {\la\phi_1|T_{np}|\phi_d\ra}.
\eeqn{DeltaE}
The representation (\ref{UNA}) coincides with the definition of the $N$--$A$ optical potential taken at energy $E_{\rm eff}$ equal to  half the deuteron incident energy plus the $n$-$p$ kinetic energy averaged over $V_{np}$ \cite{Joh14}. 
The ADWA also allows the leading order of the term 
\beq
U^{(1)} = U_{nA}\dfrac{Q_A}{e}U_{pA}+U_{pA}\dfrac{Q_A}{e}U_{nA}
\eeqn{U1}
to be evaluated. It was shown in \cite{Din19} that
\beq
\la \phi_{1}\phi_{A}|U^{(0)}&+&U^{(1)}|\phi_{A}\phi_{d}\ra \approx  2\la \phi_1 \phi_A | U^{(0)}| \phi_d \phi_A \ra
\eol
&-& \sum_{N=n,p} \la \phi_1 \phi_A | v_{NA}| \phi_d \phi_A \ra.
\eeqn{I3B}
This has a simple connection with the Nonlocal Dispersive Optical Model (NLDOM) potential given by
\beq 
\braket{\phi_1\phi_A | U^{(0)}+U^{(1)}| \phi_d\phi_A} &=& V^{HF}_{nA} + 2\Delta V_{nA}^{dyn}(E), \eol
&+& V^{HF}_{pA} + 2\Delta V_{pA}^{dyn}(E), \,\,\,\,\,\,\,\,\,\,
\eeqn{u0u1}
where $V^{HF}_{NA}$ and $\Delta V_{NA}^{dyn}(E)$ are the Hartree-Fock and dynamical part, respectively. The latter contains an imaginary part, responsible for absorption, and a real part (or polarization potential) that contains contributions from all the excited states of $A$. Both $V^{HF}_{NA}$ and $\Delta V_{NA}^{dyn}(E)$ are connected by dispersive relations \cite{mahauxsartor}. Thus, induced three-body   effects arising due to multiple scattering   in the leading order can be accounted for by a simple doubling of the dynamic part of the optical potential.

The study involving  $\la \phi_1\phi_A | U^{(0)}| \phi_d\phi_A\ra$ only revealed that including energy-dependence of the nonlocal optical potential via the energy shift, $\Delta E$, can result in a significant difference from traditional $E_d/2$-based ADWA calculations and this result strongly depends on the assumptions about the energy-dependence of the optical potentials \cite{Joh14,Wal16}. In general, applying the energy shift gives higher $(d,p)$ cross sections than those obtained in standard ADWA with local optical potentials. Including
$\la \phi_1\phi_A |U^{(1)}| \phi_d\phi_A\ra$ brings these cross sections down due to increased absorption \cite{Din19}. Interestingly, with a stronger imaginary part  increasing the dynamical real (polarisation) part does not have any noticeable consequences for $(d,p)$ cross sections. This justifies the use of phenomenological potentials that are not based on dispersive relations. This is important given that NLDOM potentials are available for a very limited number of isotopes.
Previous investigations  have been carried out mainly for the $^{40}$Ca$(d,p)^{41}$Ca reaction.  Below we will concentrate on a few $(d,p)$ reactions which are interesting from the point of view of indirect study of astrophysical $rp$-process.

\section{$(d,p)$ reactions of astrophysical interest}

Motivated by recent  and ongoing  experimental application of deuteron-induced one-nucleon transfer reactions to investigate stellar $(p,\gamma)$ reaction rates we have selected four $(d,p)$ reaction cases as indirect tools for astrophysical reaction studies. These reactions allow one to determine the
spectroscopic information, the neutron spectroscopic factor or the ANC, associated with the overlap between the wave functions of the target and residual nucleus states. Using the mirror
symmetry, discussed later in section VI, one can find the corresponding proton’s spectroscopic information needed to calculate
the proton radiative capture. The four selected reactions are the following.
\begin{itemize}
    \item $^{26}$Al$(d,p)^{27}$Al
as a surrogate for  proton capture reaction $^{26}$Al$(p,\gamma)^{27}$Si, one of the major destruction pathways of $^{26}$Al, primarily in   Wolf Rayet stars, but also in AGB stars \cite{Al_1}.
This capture reaction has been investigated with two indirect experimental  methods using $^{26}$Al beam, one measuring $^{26}$Al$(d,p)^{27}$Al angular distributions \cite{Al_dp1,Al_dp2} 
at an incident deuteron energy of $E_d=12$ MeV and another measuring total cross sections of $^{26}$Al$(d,n)^{27}$Si \cite{Al_dn} 
at $E_d=60$ MeV. 
Both experiments have been  performed to constrain the widths   $\Gamma_p$ of key $^{27}$Si resonances.
    \item $^{30}$P$(d,p)^{31}$P
as a surrogate for $^{30}$P$(p,\gamma)^{31}$S proton capture reactions, which are known to dominate the uncertainty in total abundances of sulphur isotopes \cite{P_1}, with the variation in reactions rates changing abundances by factors of up to 100 \cite{iliadis2002}. 
This uncertainty affects
the abundances of nuclei that follow later in the $rp$-process. There are no published experimental works that attempt this with $(d,p)$ reactions but measurements of total cross sections  using $(d,n)$ reactions at $E_d = 60$ MeV to constrain $^{31}$S resonances are reported in \cite{P_dn}.  
    \item $^{34}$Cl$(d,p)^{35}$Cl 
as indirect probe for $^{34}$Cl$(p,\gamma)^{35}$Ar reaction that  contributes to the production of $^{35}$Ar, an important isotopic observable in pre-solar grains in meteorites.
The importance of uncertainties associated with this reaction on calculations of final abundances in oxygen neon novae has previously been highlighted \cite{iliadis2002}, changing abundances of $^{34}$S sevenfold, 
and the ability to reduce these uncertainties is of particular importance as the $^{32}$S/$^{34}$S ratio is used to determine whether grains originate from novae \cite{Cl_1}. New states in $^{35}$Cl have been   identified in $^{36}$Cl$(d,t)^{35}$Cl reaction that include mirror analogs of astrophysical $^{35}$Ar states \cite{Fry15}, however, $(d,t)$ experiment does not allow the partial proton widths $\Gamma_p$ to be determined.

    \item  $^{56}$Ni$(d,p)^{57}$Ni.  
In the late $rp$-process and type Ia supernovae, $^{56}$Ni$(p,\gamma)^{57}$Cu reactions destroy the doubly magic $N$=$Z$ nucleus $^{56}$Ni by populating resonances in $^{57}$Cu \cite{Ni_1,Ni_2}.
Determining the production rate of the heavier than $^{56}$Ni nuclei 
relies on 
$^{57}$Cu resonance strengths and reaction rates of $^{56}$Ni$(p,\gamma)^{57}$Cu.  They have been investigated with $^{56}$Ni$(d,p)^{57}$Ni reactions using inverse kinematics, with experiments measuring differential cross sections carried out at an incident energy of $E_d=$8.9 MeV \cite{Ni_3}, with  integrated cross sections also having been measured at $E_d=64$ MeV. 
\end{itemize}
Each of these reactions will be investigated for 
a number of excited states, with a preference for states where experimental cross sections are available. The excitation energies and spin parities of selected states probed for each reaction are listed in Table \Ref{states_table}.


 All ADWA $(d,p)$ calculations presented in this paper have been performed using the {\sc{twofnr}} code \cite{twofnr}, which has an option of reading externally generated distorted waves in and thus can easily accommodate any calculations with nonlocal potentials.
 When using nonlocal nucleon potentials  in deuteron and proton channels we calculate the corresponding distorted waves as explained in \cite{Bai17}. 
All cross sections presented below are calculated in the zero-range approximation for $V_{np}(\ve{r})\phi_d (\ve{r})\approx D_0 \delta(\ve{r})$ in the $(d,p)$ reaction $T$-matrix, which is consistent with assumptions made in \cite{Joh14,Din19} to derive expressions  (\ref{UNA}),  (\ref{U1}) and (\ref{I3B}). $D_0$ value does not depend on NN model choice, so we use a standard value of $D_0=-126.15$ MeV fm$^{3/2}$.
The calculations are carried out with overlap integrals represented by single-particle wave functions calculated in a Woods-Saxon potential well with a radius $r_0=1.25$ fm, diffuseness $a=0.65$ fm and spin orbit depth $V_{s.o.}=6$ MeV. For the purposes of comparing different model calculations we employ the spectroscopic factor $S$ equal to one.

\section{Numerical results for angular distributions}

\subsection{Cross sections with different energy shift defined by deuteron model choice}

\newcolumntype{C}[1]{>{\centering\arraybackslash}p{#1}}
\begin{table}[b!]
\begin{center}
 \caption{$\Delta E$ values (in MeV) and percentage of $d$-state contribution for the different deuteron wavefunction models used.}   \begin{tabular}{C{4cm}|C{2cm}|C{2cm}}
        \hline
        Wavefunction model & $\Delta E$   & D-state contribution \\
        \hline
	     Hulth\'en \cite{Hulthen}           & 57    & 0 \% \\
	     AV18 \cite{AV18}                   & 63.1  & 5.76 \% \\
	     CDBonn \cite{CDBonn}               & 56.3  & 4.85 \% \\
	     $\chi$EFT at N4LO with regulator of
	    0.8 fm \cite{N4LO} & 123.6 & 4.29 \% \\
	     \hline
    \end{tabular}   
    \label{Delta_E_table}
\end{center}{}
\end{table}

\begin{figure*}[t!]
\centering
\includegraphics[scale=0.24394]{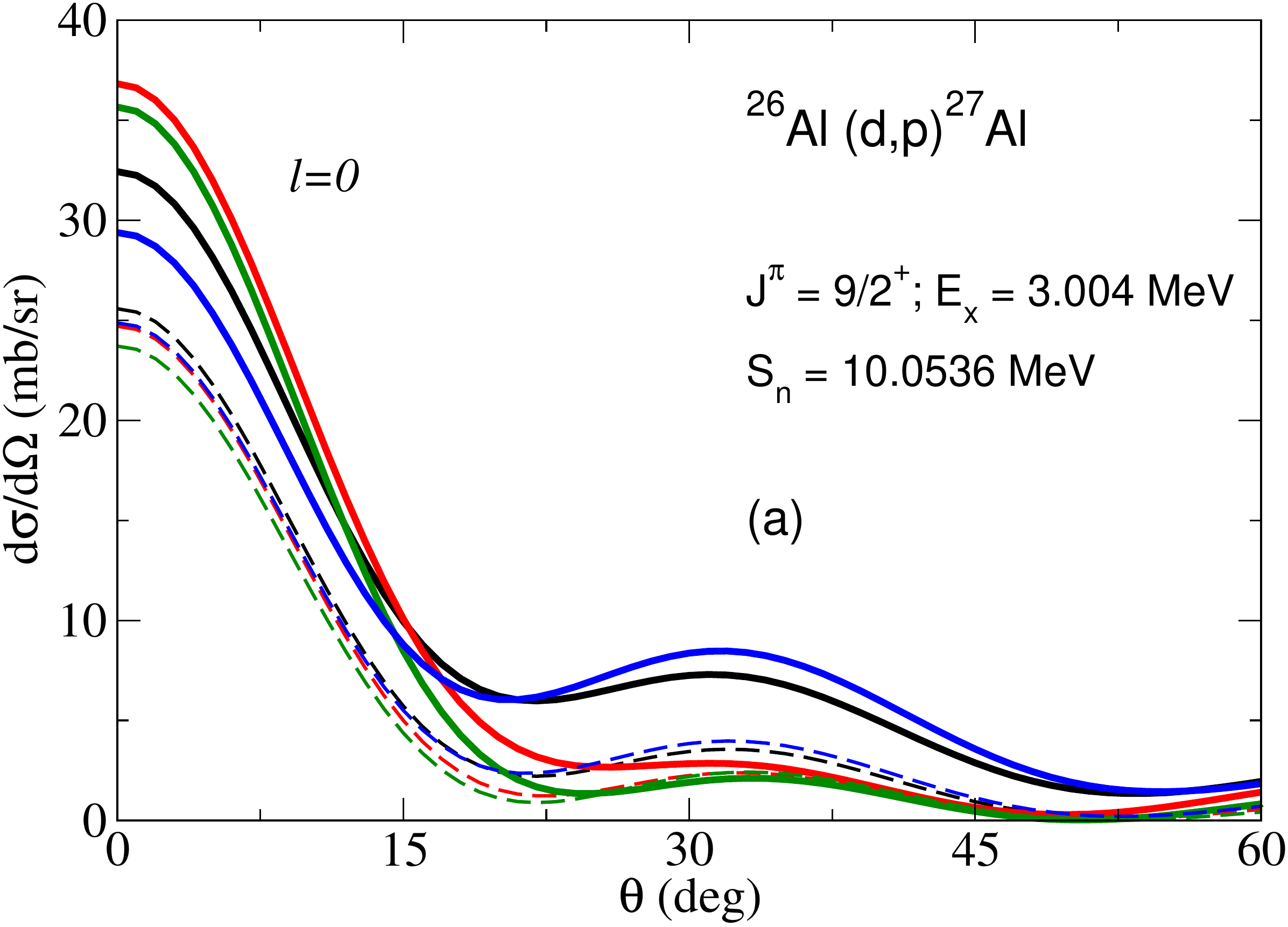}
\includegraphics[scale=0.24394]{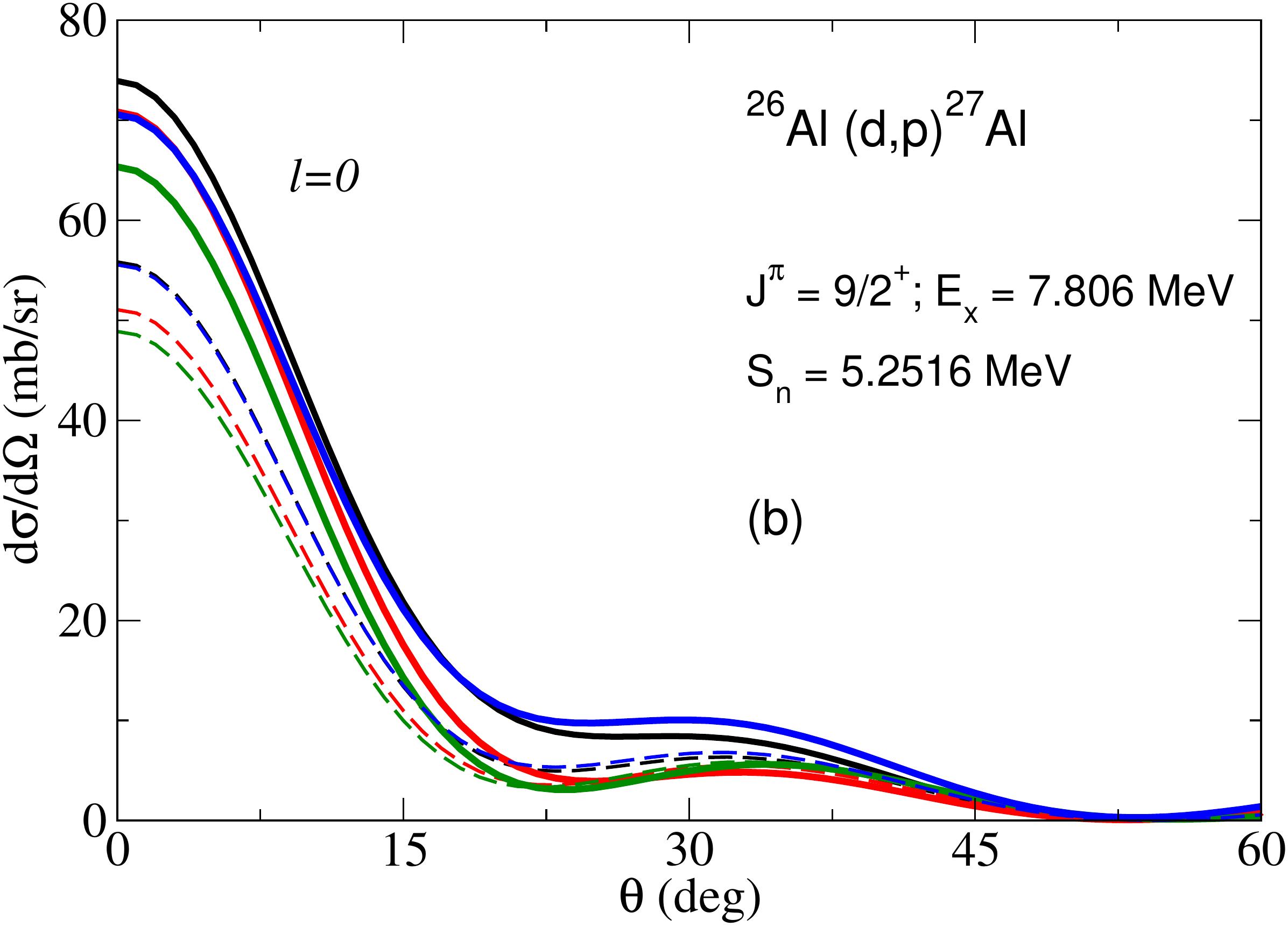}
\includegraphics[scale=0.24394]{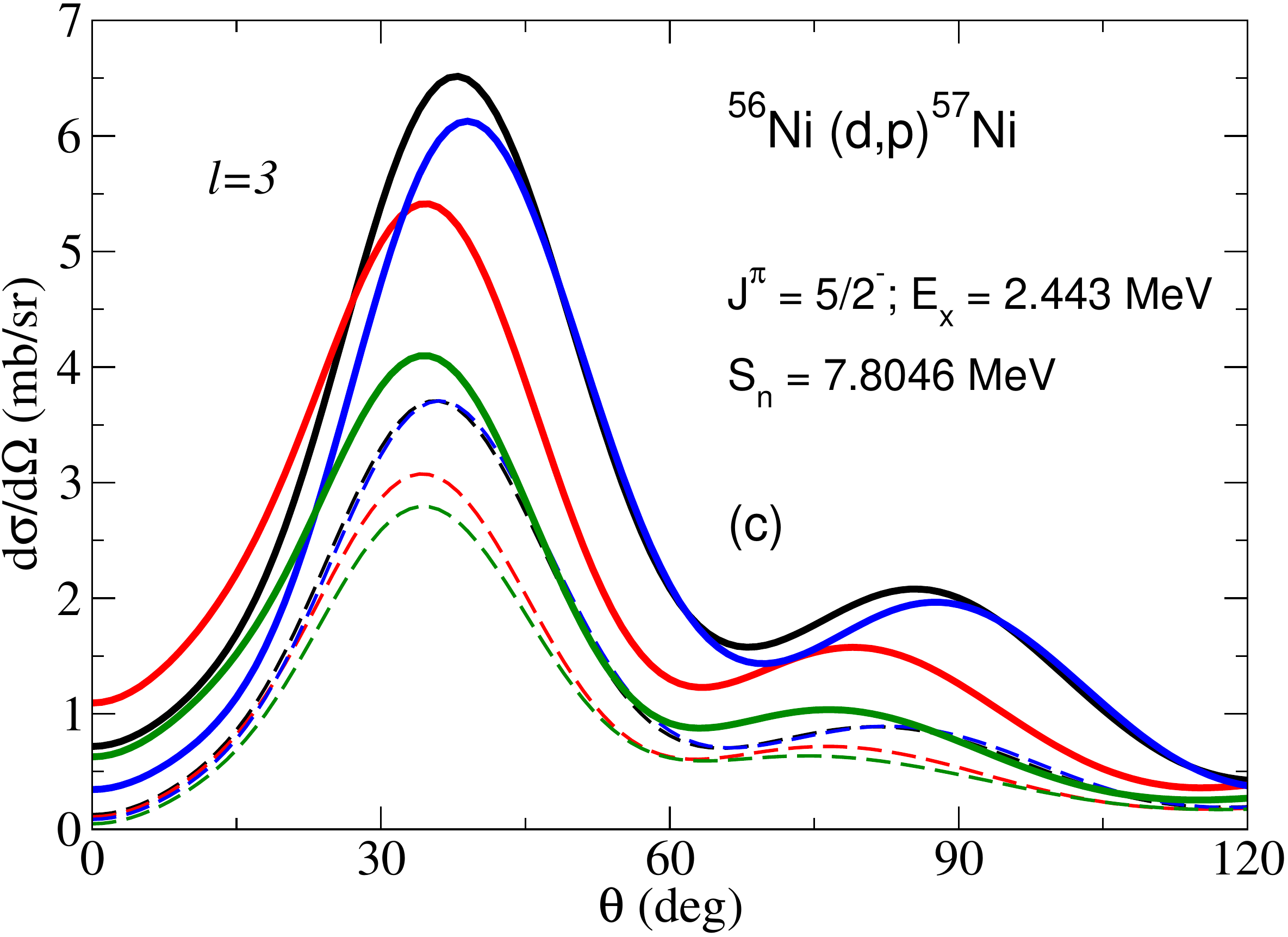}
\includegraphics[scale=0.24394]{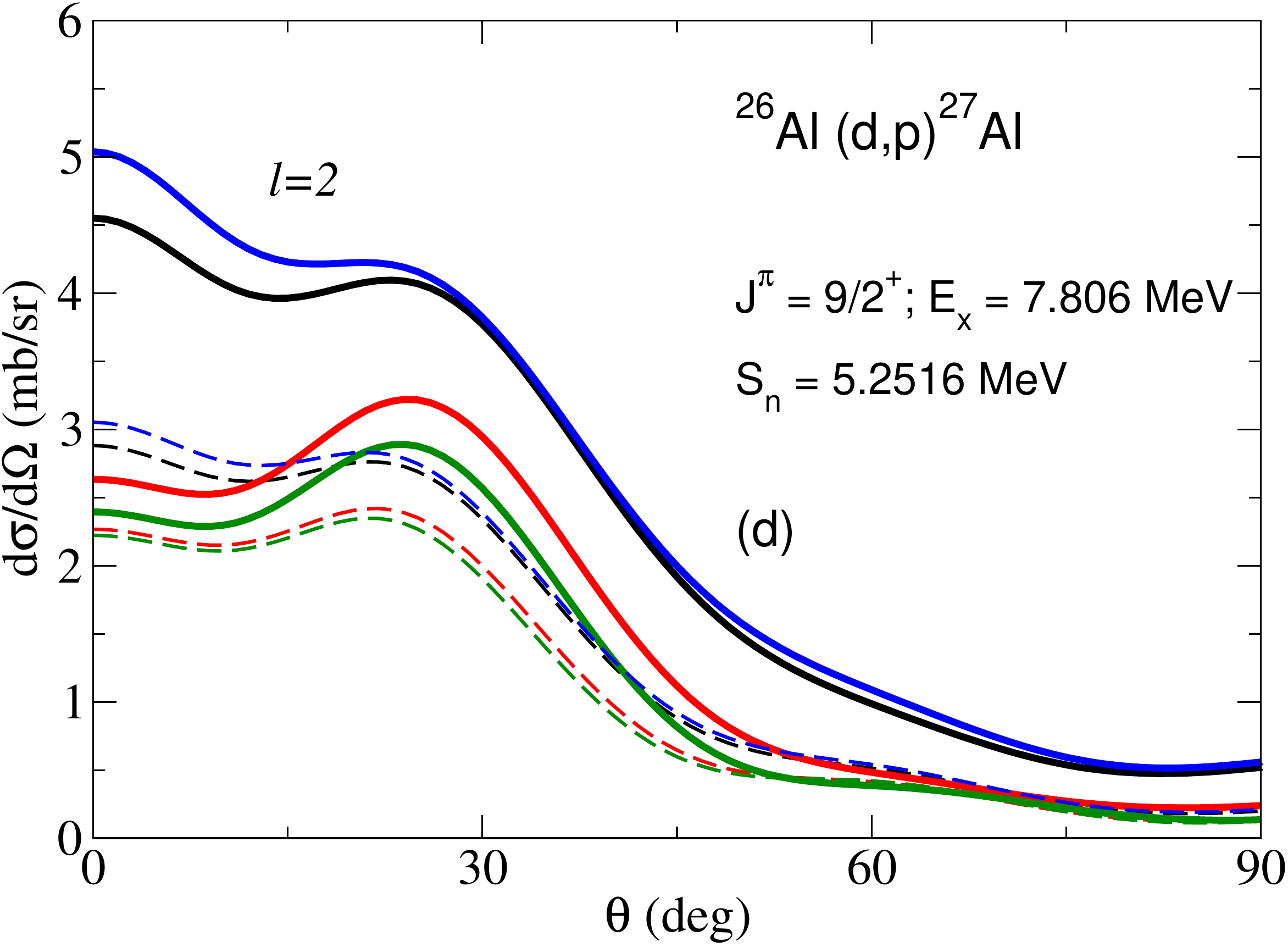}
\includegraphics[scale=0.24394]{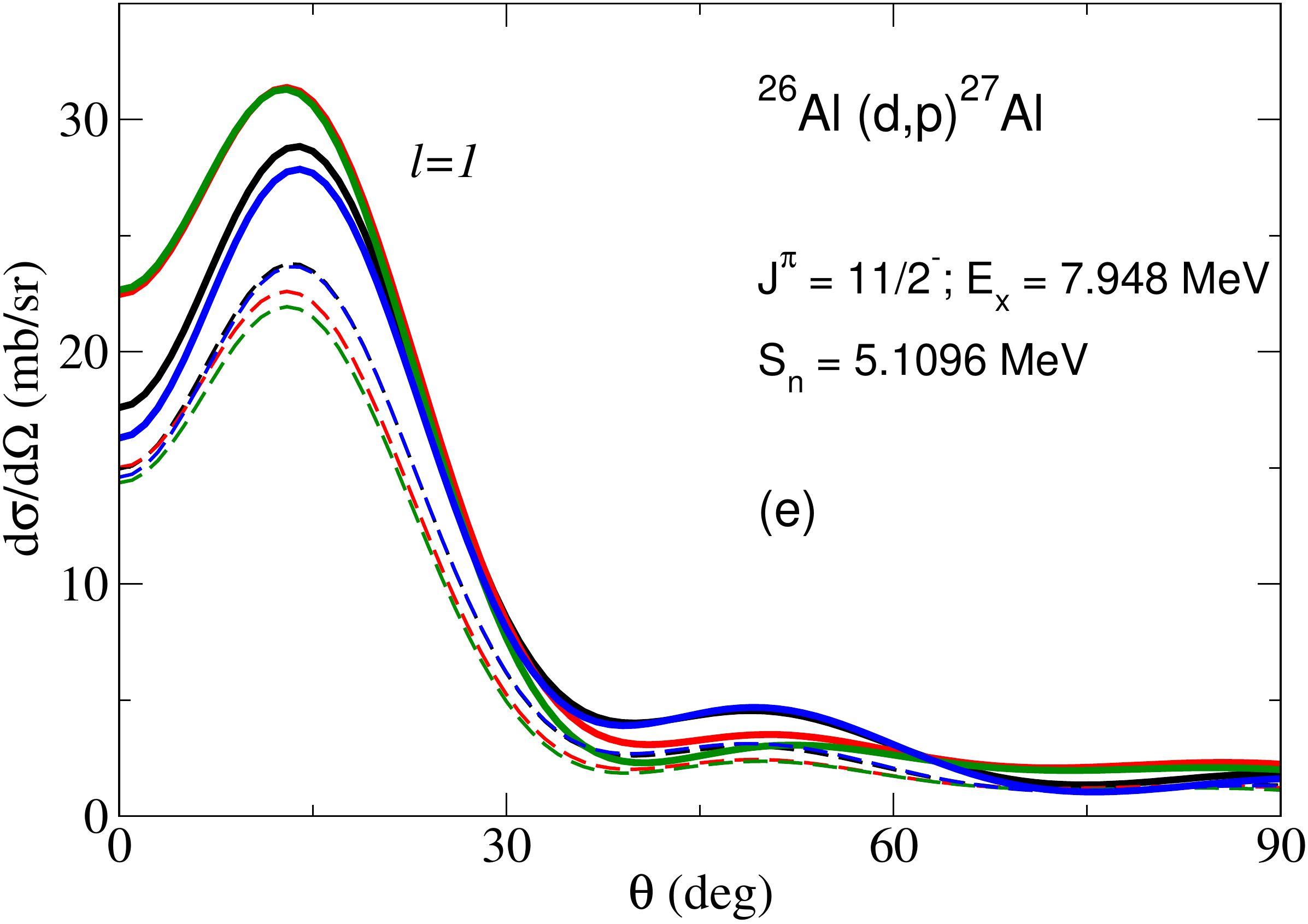}
\includegraphics[scale=0.24394]{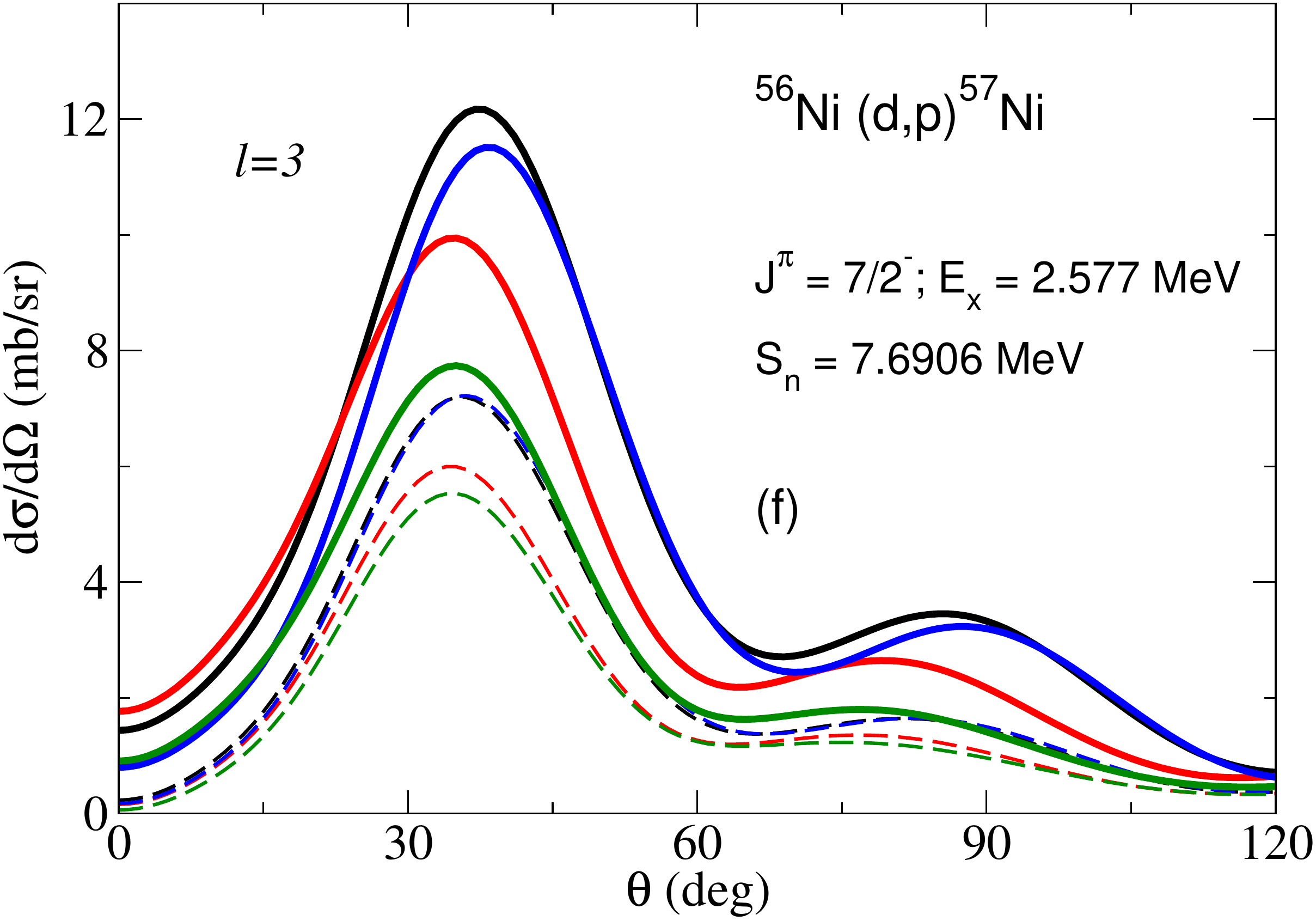}
\caption{Comparison of $^{26}$Al$(d,p)^{27}$Al $(a,b,d,e)$ and  $^{56}$Ni$(d,p)^{57}$Ni $(c,f)$ cross sections calculated at $E_{d}$ = 12 MeV with GRZ potentials  with (dashed lines) and without (solid lines) I3B terms for NN models from Table \ref{Delta_E_table}. The results with Hulth\'en, AV18, CD-Bonn and $\chi$EFT NN model choices are shown by green, black, red and blue lines, respectively.
}
\label{NL_cs}
\end{figure*}

We start with presenting calculations using global systematics of nonlocal optical potentials from Giannini, Ricco and Zucchiatti (GRZ) \cite{GRZ}. This potential has energy-independent real part and energy-dependent imaginary part, which vanished at zero nucleon scattering energy and approaches exponentially to a constant value of 17.5 MeV with its increase. 
For this potential employing the shift  $\Delta E$ is crucial since without it the imaginary part is too small to give sensible cross sections \cite{Joh14}.
However, the numerical value of $\Delta E$ strongly depends on the choice of the NN interaction model, varying from 57 to 123 MeV \cite{Bai16}. Some of these values are shown in Table \ref{Delta_E_table}.
The $n$-$p$ model choice for $V_{np}$ in Eq. (\ref{H3}) also affects the ADWA solution for $d$-$A$ distorted wave  \cite{Bai16} so that we have two sources of NN-model-induced uncertainties when using energy-dependent nonlocal potentials.

To disentangle these uncertainties we first fix $V_{np}$ potential in the three-body Schr\"odinger equation choosing the Hulth\'en model \cite{Hulthen}, which results in using the Hulth\'en wave functions for $\phi_d$ and $\phi_1$ in Eq. (\ref{2bAD}), and make a series of calculations with GRZ potentials shifted by energies from 0 to 100 MeV with a step of 20 MeV. We have chosen $^{26}$Al($d,p)^{27}$Al$^*$($E_x=7806$ keV)  at $E_d = 12$ MeV as an example and studied it both for $l=0$ and $l=2$ orbital momentum transfer. Similar to what has been observed in \cite{Joh14}, $\Delta E = 0$ leads to significantly higher cross sections in the main peak, however, for $\Delta E$ between 57 to 100 MeV,  spanning most values generated by different NN models, the corresponding difference in angular distributions are small. This is explained by the fact that imaginary part of GRZ for $E > 60$ MeV is practically saturated and its energy-dependence is very weak. Doubling the imaginary part according to prescription of Eq. (\ref{I3B})  does not change this dependence.

\begin{figure*}[t]
\includegraphics[scale=0.2433]{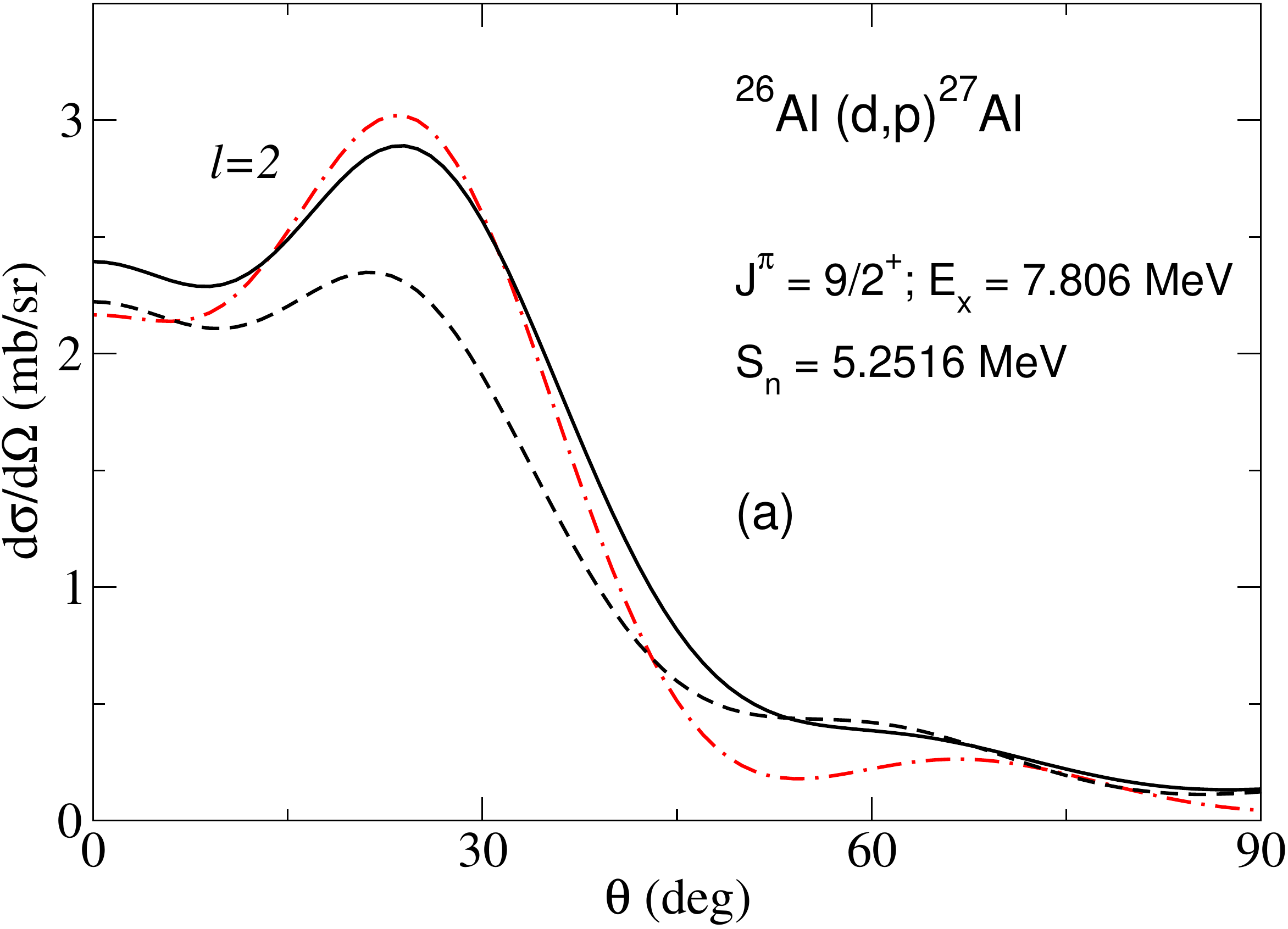}
\includegraphics[scale=0.2433]{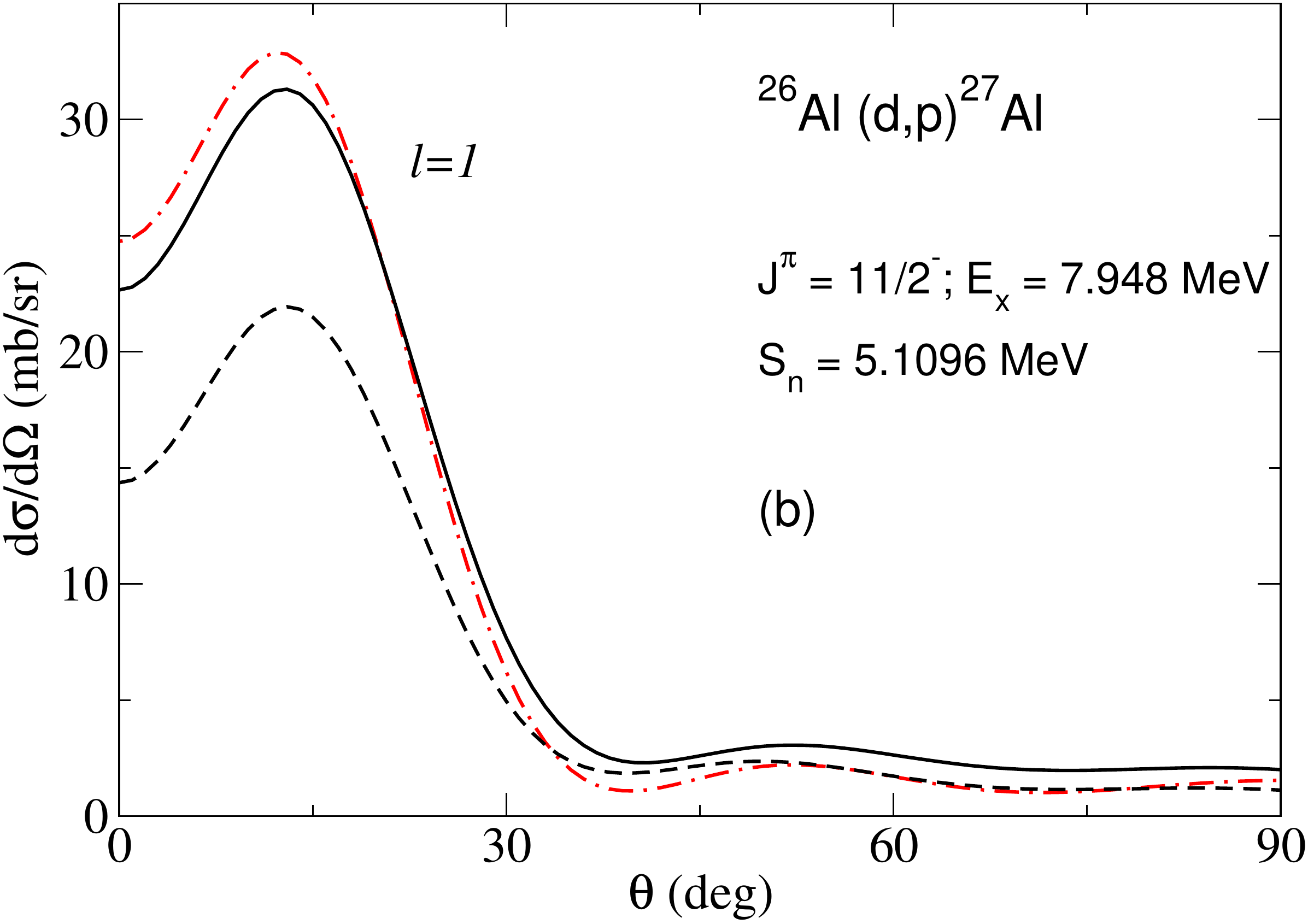}
\includegraphics[scale=0.2433]{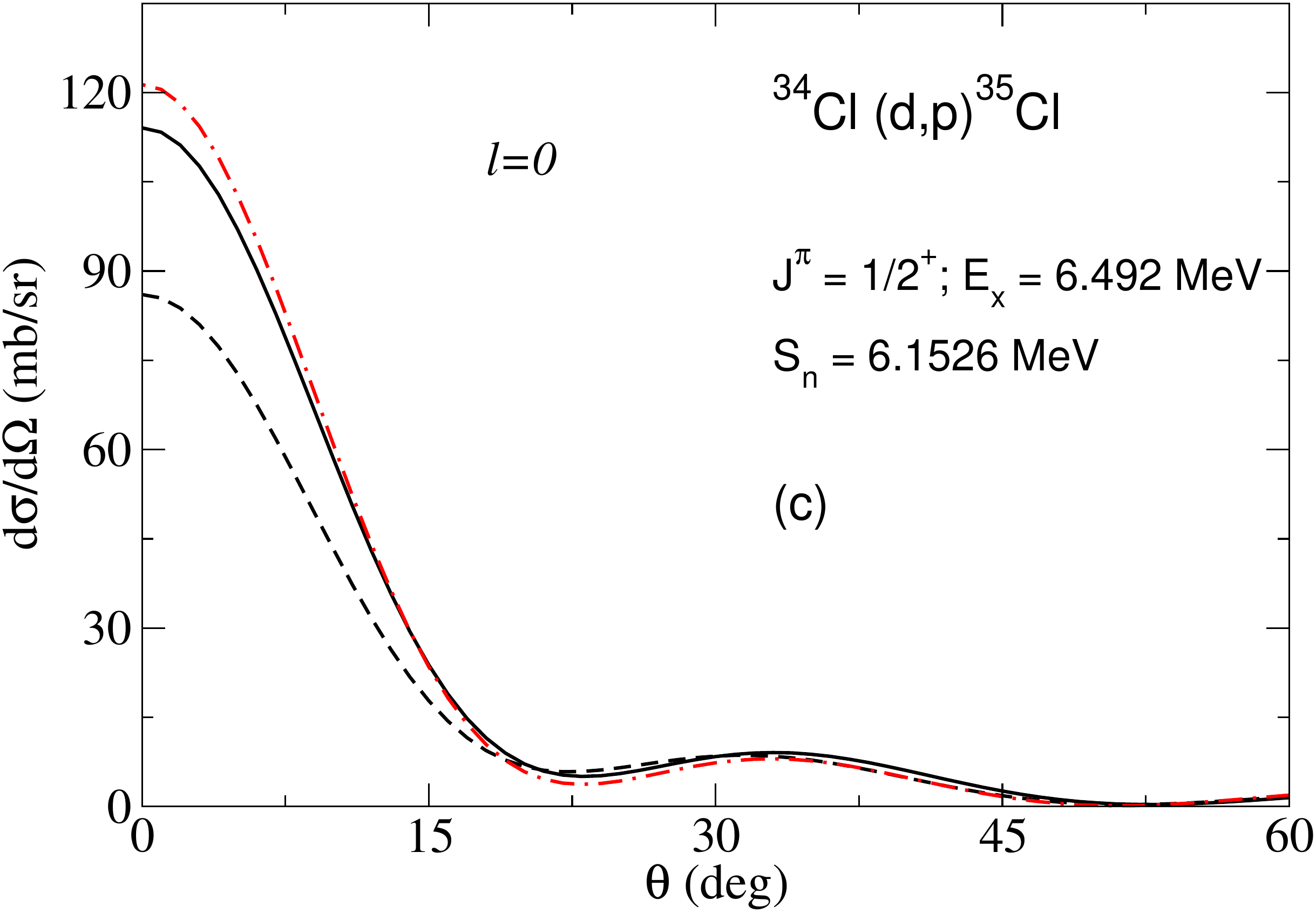}
\includegraphics[scale=0.2433]{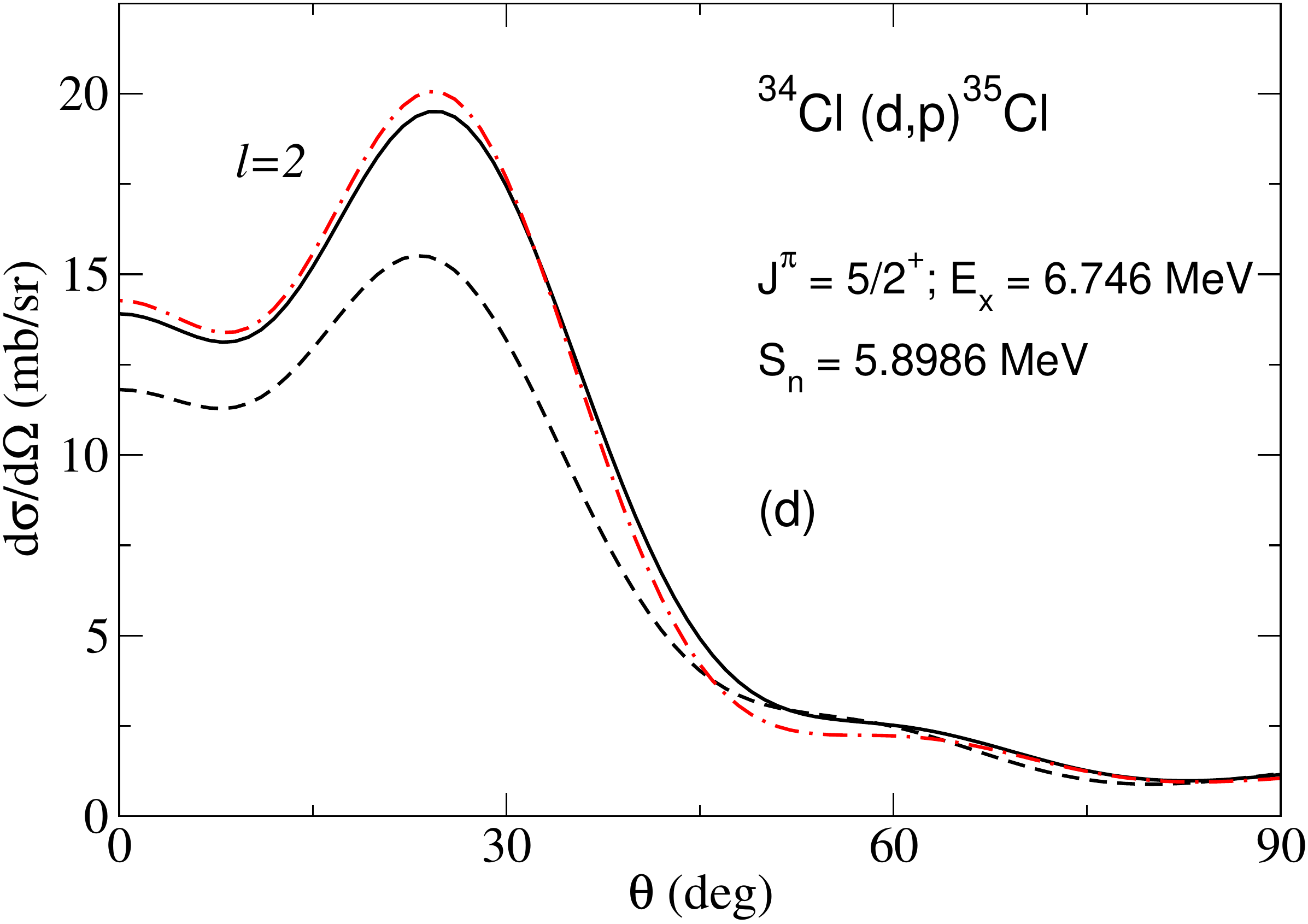}
\includegraphics[scale=0.2433]{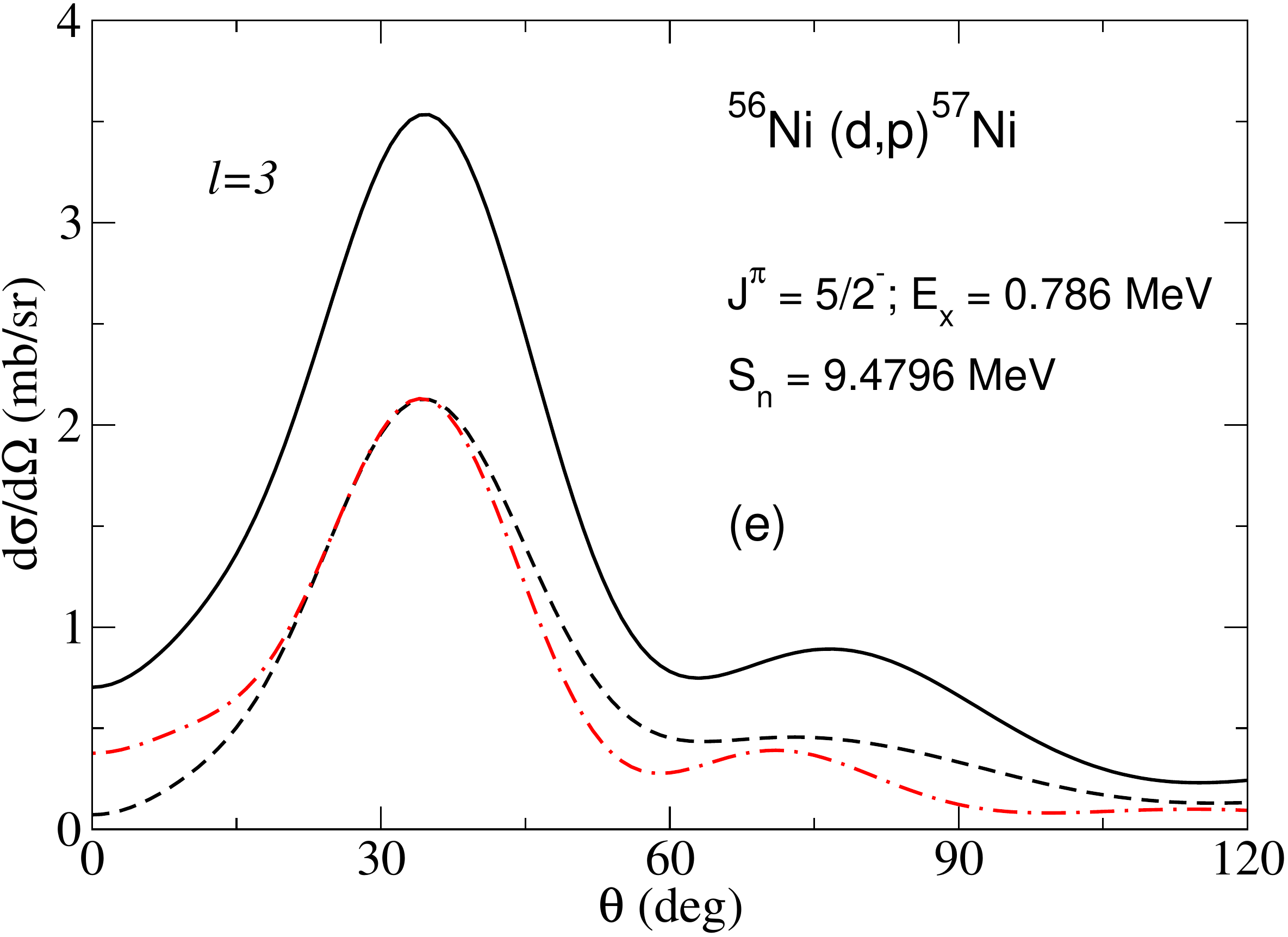}
\includegraphics[scale=0.2433]{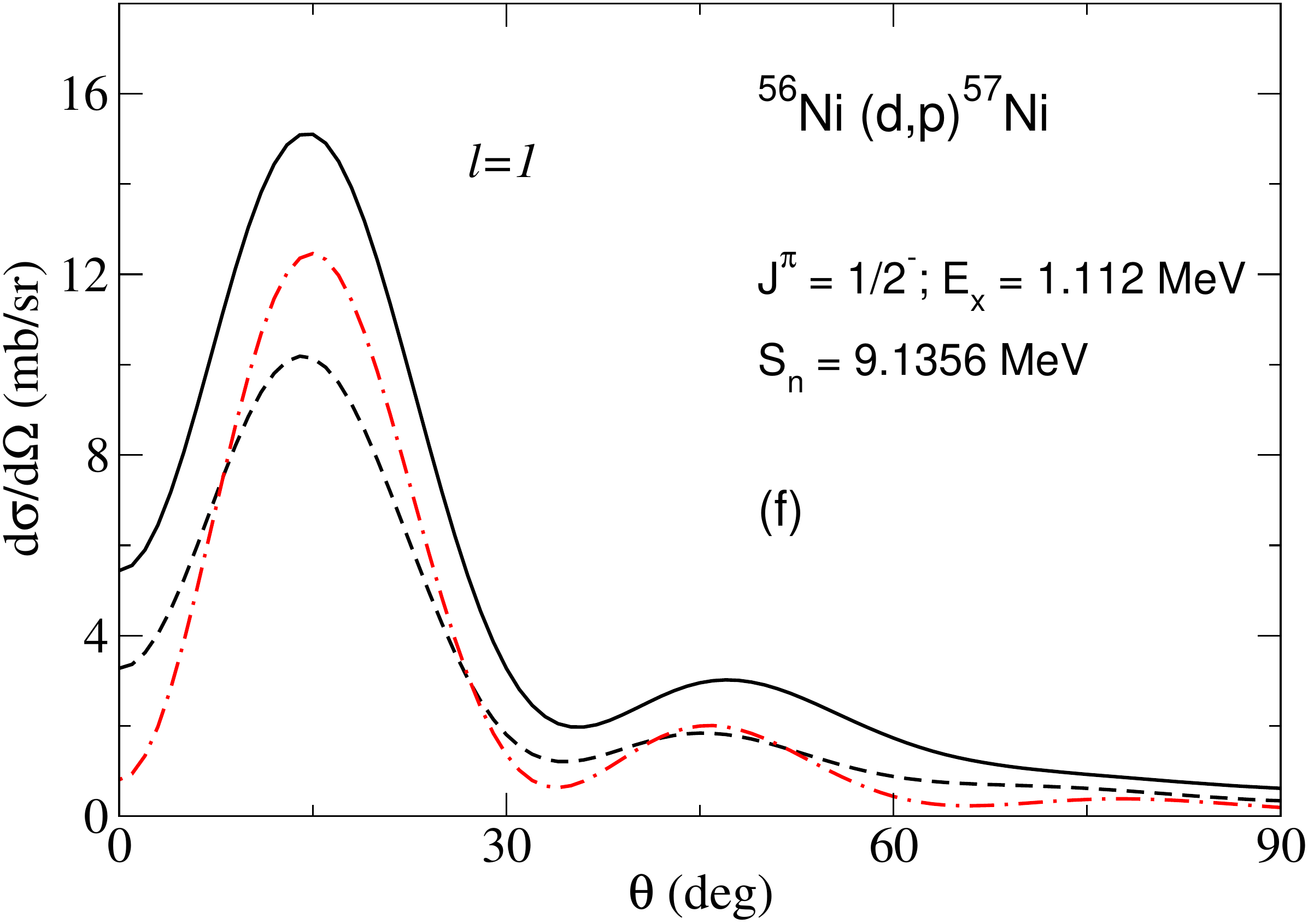}
\caption{The ADWA  angular distributions for  $^{26}$Al$(d,p)$ $(a,b)$,  $^{34}$Cl$(d,p)$ $(c,d)$ and  $^{56}$Ni$(d,p)$ $(e,f)$ reactions and populating $l=0$ ($c$), $l=1$ ($b,f)$, $l=2$ ($a,d$) and $l=3$ ($e$) states calculated  at $E_d=$ 12 MeV  using the nonlocal GRZ potential with (dashed) and without (solid) I3B terms in comparison with those obtained using standard Johnson-Tandy model with the local KD03 potential (dot-dashed). }
\label{local_GRZ_KD_comp_figure2}
\end{figure*}

As a next step, we used consistently the same NN model both in evaluating the shift $\Delta E$ and the functions $\phi_d$ and $\phi_1$ that enter the matrix element
$\la \phi_1|U^{(0)} + U^{(1)} | \phi_d\ra$. We used four NN models, presented in Table \ref{Delta_E_table}. The first of them, the Hulth\'en model, contains deuteron $s$-wave state only. The other three models include contribution from deuteron $d$-wave state. We  account for this contribution exactly using approach developed in \cite{Bai17}. In Fig. \ref{NL_cs} we show  angular distributions for two $(d,p)$ reactions populating  mirror analogs of astrophysically relevant states, $^{26}$Al$(d,p)^{27}$Al and $^{56}$Ni$(d,p)^{57}$Ni.
The calculations are performed without and with doubling imaginary part of GRZ,  which corresponds to the absence and presence of the I3B  effects, respectively. 
The figure shows that without I3B effects the spread in the angular distributions due to different NN model choice is significant, up to 25\% with respect to the average value for $l=2$ and $l=3$ transfers. This is similar to findings of Ref. \cite{Bai16} where energy-independent nonlocal potential was used. However, for $l=0$  and $l=1$ cases the spread obtained with GRZ, around 5-15\% and although being smaller than that for $l=2,3$,   is larger than that obtained in \cite{Bai16}. Also, in the case of the energy-independent nonlocal potential of \cite{Bai16} the cross sections in the main peak are clearly correlated with the matrix element $\la \phi_1 | T_{np} | \phi_d\ra$, being larger with increased values of the latter. This correlation seems to be either lost or not so well expressed for the energy-dependent potential GRZ when in some cases the cross sections for the smallest  $\la \phi_1 | T_{np} | \phi_d\ra$ are the largest. This can be a result of small changes in the optical potential evaluated for different $\Delta E$, when a smaller  $\la \phi_1 | T_{np} | \phi_d\ra$ leads to a smaller imaginary part, which produces less absorption. When the I3B force is added, the spread between the cross sections obtained with different NN models is significantly reduced for all cases considered. It does not exceed 15\% with respect to the average value for $l=2,3$ transfers, and is no larger than 10\% for $l=0$ and $l=1$ transfers. Also, the correlation between   $\la \phi_1 | T_{np} | \phi_d\ra$ is  restored. Similar to findings of \cite{Din19}, the introduction of I3B significantly lowers the cross sections.
We have performed calculations (not shown here) for two other reactions including $sd$-shell targets $^{30}$P and $^{34}$Cl at the same deuteron incident energy of 12 MeV and populating the $^{31}$P and $^{35}$Cl states from Table \ref{states_table}. We came to the same conclusions about I3B effects and NN model choice as in the case of $^{26}$Al and $^{56}$Ni.  

\newcolumntype{C}[1]{>{\centering\arraybackslash}p{#1}}
\begin{table*}[htb]
\begin{center}
\caption{The ratio between the maximum of the differential cross sections peaks calculated at $E_{d}$ = 12 MeV for reactions from the first column. The excitation energies (in keV), final state spins and quantum numbers of the populated level are given in the second, third and fourth column. The position of the maximum, $\theta$, are given in the fifth column. In the following columns $\sigma_{\text{GRZ}}$ and  $\sigma^{\rm I3B}_{\text{GRZ}}$ denote  the cross sections calculated using GRZ  without and with I3B, respectively, $\sigma_{\rm KD03}$ are standard Johnson-Tandy results with KD03 potentials, $\sigma_{\rm KD03}^{\rm I3B}$ are obtained by applying shift $\Delta E$ and I3B effects to the local potential KD03 and $\sigma_{\rm KD03}^{\rm NLE,I3B}$ is obtained by restoring nonlocal equivalent of the local KD03 potential.}

\begin{tabular}{C{2.2cm}|C{1cm}C{1cm}C{1cm}|C{1cm}C{2cm}C{2cm}C{2cm}C{2cm}C{2cm}}
 \hline
  & & & & & & & & & \\
 Reaction & $E_x$ & $J^{\pi}$ & $lj$ & $\theta$ & $\sigma_{\text{GRZ}}/\sigma_{\rm KD03}$ & $\sigma^{\rm I3B}_{\text{GRZ}}/\sigma_{\rm KD03}$ & $\sigma^{\rm I3B}_{\text{GRZ}}/\sigma_{\rm GRZ}$ & $\sigma^{\rm I3B}_{\text{KD03}}/\sigma_{\rm KD03}$ & $\sigma^{\rm NLE,I3B}_{\text{KD03}}/\sigma_{\rm KD03}$ \\ 
  & & & & & & & & & \\
 \hline 
 $^{26}$Al$(d,p)^{27}$Al & 3004 & 9/2$^{+}$ & $s_{\tfrac{1}{2}}$ & 0$^\circ$  & 0.924 & 0.614 & 0.665 & 0.896 & 0.868 \\
                         & 7806 & 9/2$^{+}$ & $s_{\tfrac{1}{2}}$ & 0$^\circ$  & 0.945 & 0.707 & 0.748 & 1.041 & 1.030 \\
                         & 7806 & 9/2$^{+}$ & $d_{\tfrac{3}{2}}$ & 22$^\circ$ & 0.956 & 0.782 & 0.818 & 0.962 & 0.970 \\
                         & 7948 & 11/2$^{-}$& $p_{\tfrac{1}{2}}$ & 13$^\circ$ & 0.954 & 0.688 & 0.701 & 0.827 & 0.825 \\
 \hline
 $^{30}$P$(d,p)^{31}$P   & 6336 & 1/2$^{+}$ & $s_{\tfrac{1}{2}}$ & 0$^\circ$  & 0.989 & 0.736 & 0.744 & 1.046 & 1.036 \\
                         & 6336 & 1/2$^{+}$ & $d_{\tfrac{3}{2}}$ & 24$^\circ$ & 1.018 & 0.807 & 0.793 & 0.992 & 0.997 \\
                         & 6399 & 7/2$^{-}$ & $f_{\tfrac{5}{2}}$ & 35$^\circ$ & 1.208 & 0.812 & 0.673 & 0.907 & 0.907 \\
 \hline
 $^{34}$Cl$(d,p)^{35}$Cl & 6492 & 1/2$^{+}$ & $s_{\tfrac{1}{2}}$ & 0$^\circ$  & 0.940 & 0.709 & 0.754 & 1.008 & 1.006 \\
                         & 6746 & 5/2$^{+}$ & $d_{\tfrac{5}{2}}$ & 24$^\circ$ & 0.972 & 0.772 & 0.794 & 0.921 & 0.939 \\
 \hline
 $^{56}$Ni$(d,p)^{57}$Ni & 768  & 5/2$^{-}$ & $f_{\tfrac{5}{2}}$ & 34$^\circ$ & 1.659 & 0.998 & 0.602 & 1.154 & 1.106 \\
                         & 1112 & 1/2$^{-}$ & $p_{\tfrac{1}{2}}$ & 15$^\circ$ & 1.212 & 0.814 & 0.672 & 1.015 & 0.993 \\
                         & 2443 & 5/2$^{-}$ & $f_{\tfrac{5}{2}}$ & 35$^\circ$ & 1.472 & 1.003 & 0.681 & 1.111 & 1.089 \\
                         & 2577 & 7/2$^{-}$ & $f_{\tfrac{7}{2}}$ & 35$^\circ$ & 1.255 & 0.897 & 0.715 & 0.961 & 0.963 \\
 \hline
 \end{tabular}   
 \label{states_table}
\end{center}{}
\end{table*}




\subsection{Nonlocal optical potential and deuteron $s$-wave model }

The strong dependence of $\Delta E$ on NN model arises in ADWA mainly due to including a deuteron $d$-state in $\phi_1$ \cite{Bai16,Bai17}, which opens the door to a large contribution from  poorly-known high $n$-$p$ momenta  into $\la \phi_1 | T_{np} | \phi_d \ra$. Such a contribution also affects the ADWA $d$-$A$ scattering waves and $(d,p)$ cross sections, obtained with nonlocal optical potentials, which
 is a drawback of ADWA  that should disappear when three-body $A+n+p$ dynamics is treated exactly \cite{Del18,Gom18}.
However, if only the $s$-wave state is retained in the deuteron wave function then the contribution to $\Delta E$ is dominated by low $n$-$p$ momentum physics where sensitivity to the $n$-$p$ model is weak. In the rest of the paper we choose the Hulth\'en model for $V_{np}$, $\phi_d$, $\phi_1$ and $\Delta E$. 
As shown in  \cite{Gom18}, in this case ADWA with nonlocal energy-independent potentials gives  $(d,p)$ results that are closer to  those obtained in exact three-body calculations.

We have calculated angular distributions for reactions from  Table \ref{states_table} using the GRZ potential evaluated for $\Delta E= 57$ MeV  without and with  I3B force. We compare them to the standard Johnson-Tandy ADWA cross sections obtained with local optical potential KD03 \cite{KD03}. These calculations have been carried out for an incident deuteron energy of 12 MeV available at several radioactive beam facilities. We have discovered that without the I3B force  the main peak cross sections $\sigma_{\rm KD03}$ and $\sigma_{\rm GRZ}$, obtained with GRZ  and KD03, respectively,  are very similar for populating  $^{27}$Al, $^{31}$P$^*(6336)$ and $^{35}$Cl excited states. Their ratio is shown in Table \ref{states_table}. In the case of $^{31}$P$^*(6336)$ and $^{57}$Ni final states, $\sigma_{\rm GRZ}$ is larger than $\sigma_{\rm KD03}$ by 20-66\%. A few selected angular distributions for populating $1s$, $0d$, $1p$ and $0f$ states are shown in Fig. \ref{local_GRZ_KD_comp_figure2}. One can see that including an energy shift does not change the shape of angular distributions very much, resulting mainly in the renormalization of the cross sections.
Including I3B force leads to decrease of the peak cross section by 60-80\% as shown in Table \ref{states_table} by ratios $\sigma_{\rm GRZ}^{\rm I3B}/\sigma_{\rm GRZ}$. As the result, in most of the peak cross sections with GRZ and I3B force are much smaller than those obtained with KD03, only a few cases being within 10\% of the KD03 results.


The ratios $\sigma_{\rm GRZ}^{\rm I3B}/\sigma_{\rm KD03}$ indicate how much the spectroscopic factors change if the GRZ+I3B model is used instead of the ADWA with KD03. If these spectroscopic factors are used for $(p,\gamma)$ determination then the corresponding model uncertainty will propagate to the proton capture cross sections in similar proportions.

\subsection{Local global systematics}

Most global systematics of optical potentials are local. It is therefore of interest to check the consequences of applying an adiabatic model of three-body optical potentials for the case of local potentials. We have chosen the KD03 systematics for $n$-$A$ and $p$-$A$ potentials, which were evaluated at $E_{\rm eff} = E_d/2+\Delta E$ with $\Delta E = 57$ MeV  to account for energy-dependence of optical potentials and then multiplied their imaginary parts by the factor of two. In order to gauge the impact of these modifications to $N$-$A$ potentials we compare the results obtained with those from the standard Johnson-Tandy ADWA approach employing the same  KD03 potential but without any modifications.

\begin{figure*}[htb]
\includegraphics[scale=0.2424]{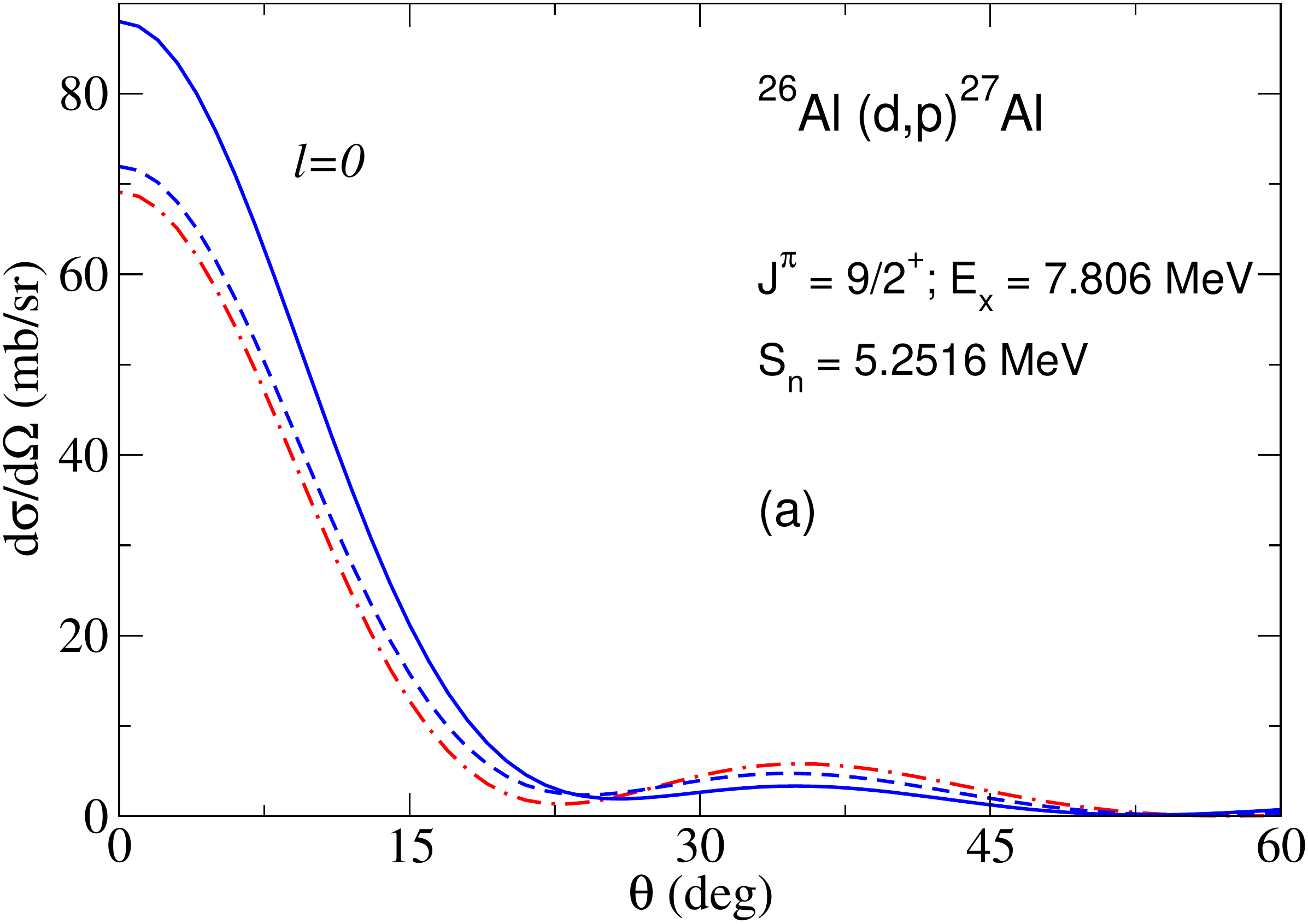}
\includegraphics[scale=0.2424]{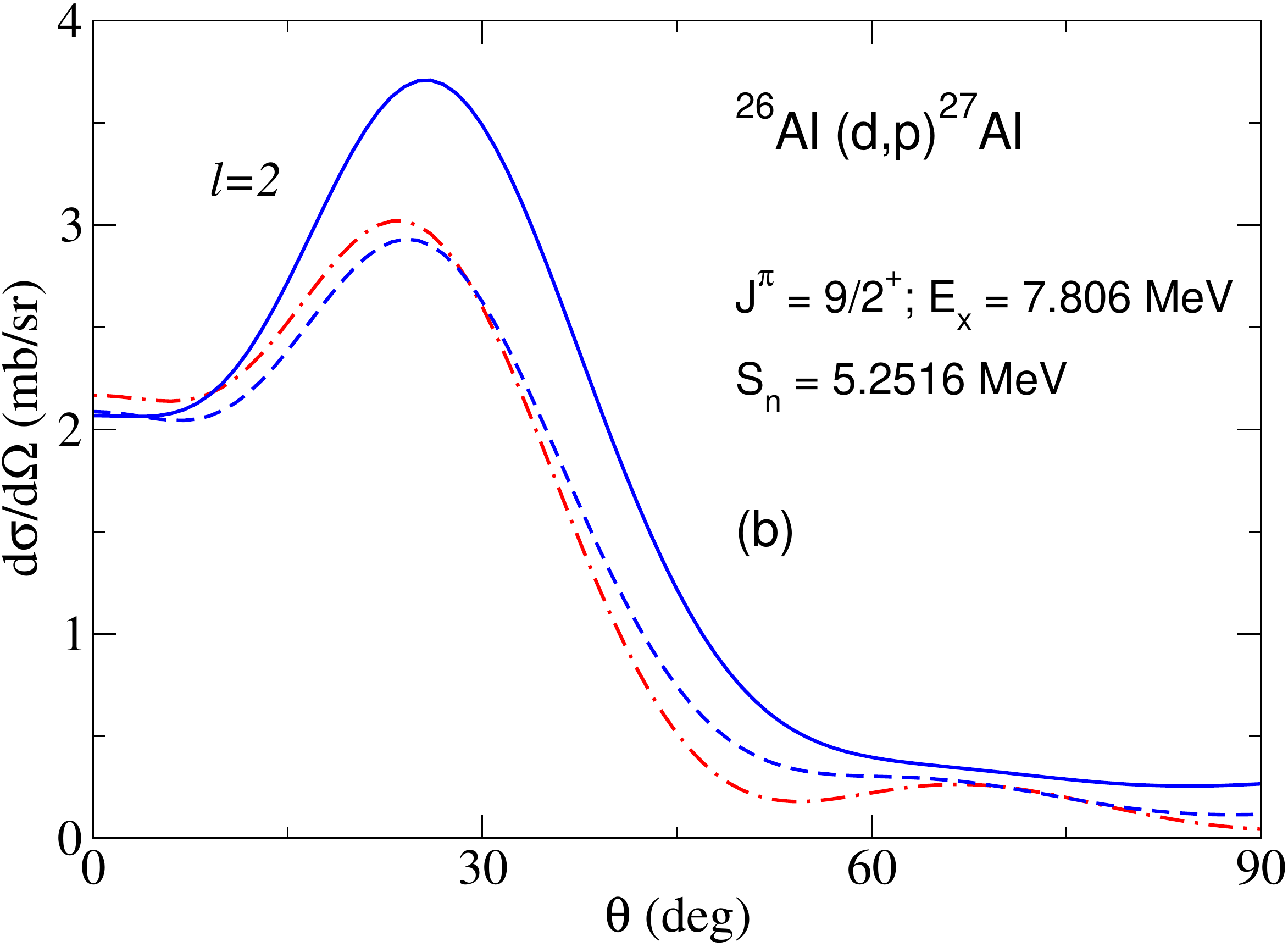}
\includegraphics[scale=0.2424]{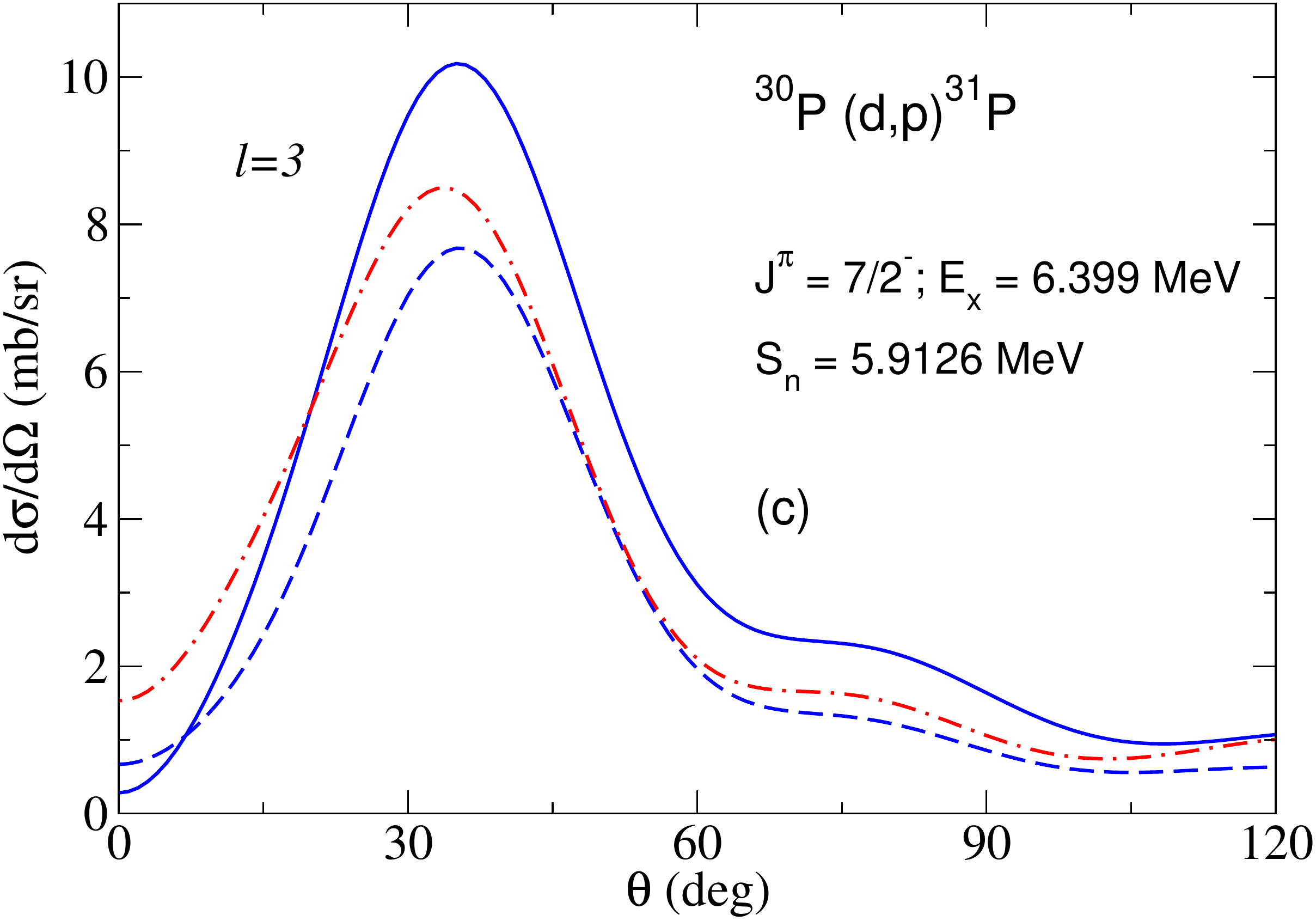}
\includegraphics[scale=0.2424]{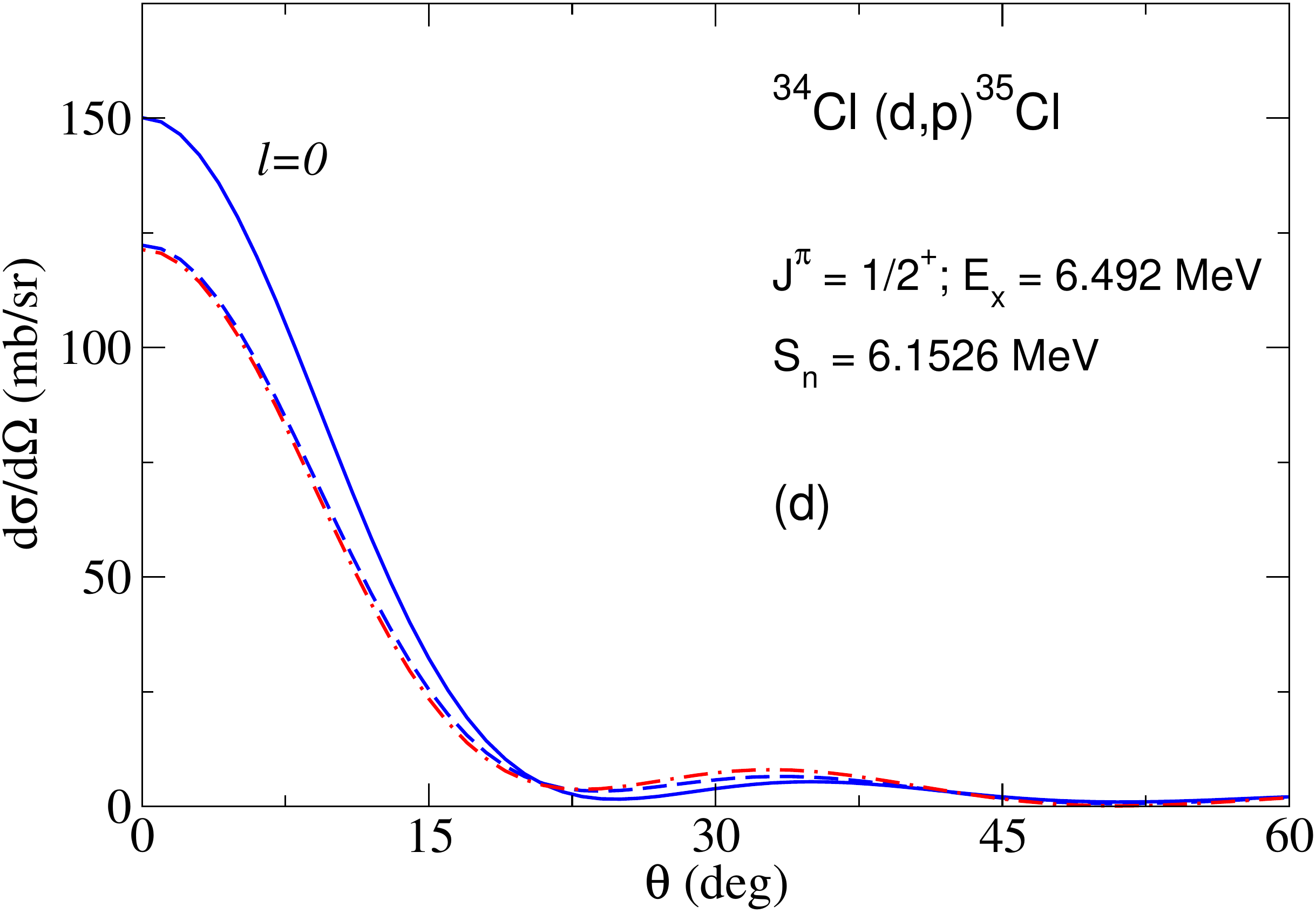}
\includegraphics[scale=0.2424]{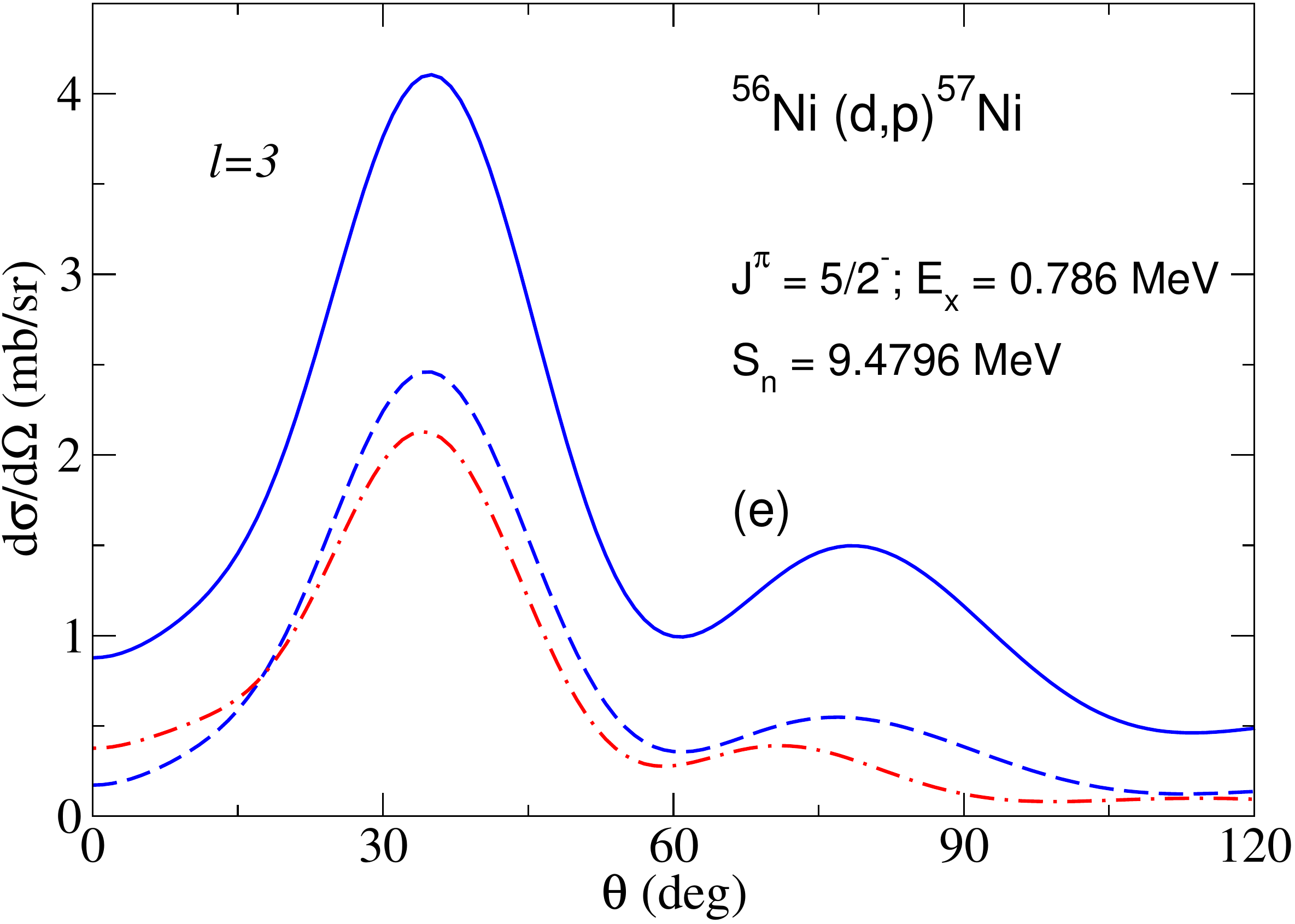}
\includegraphics[scale=0.2424]{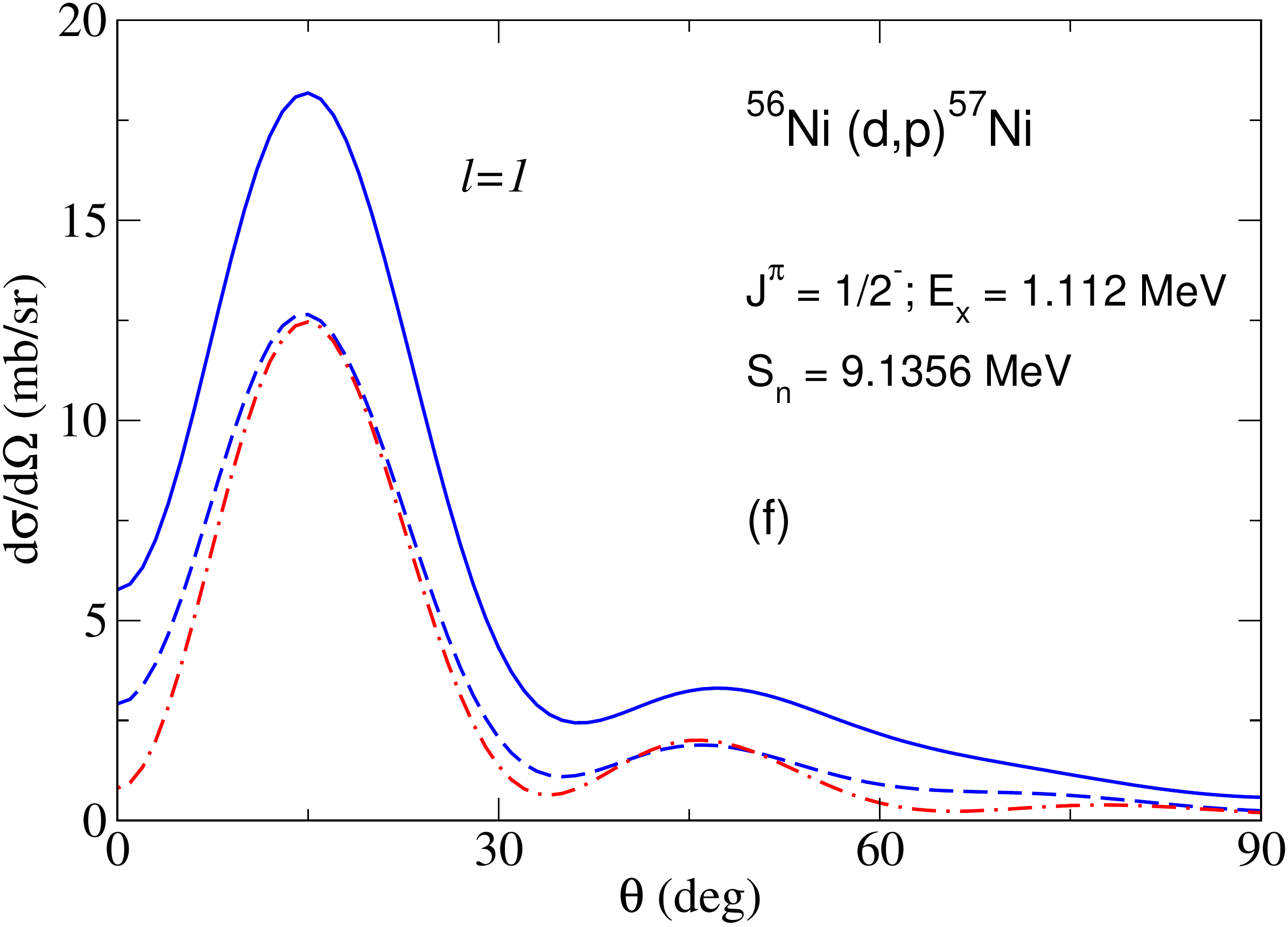}
\caption{
The ADWA  angular distributions for  $^{26}$Al$(d,p)$ $(a,b)$,  $^{30}$P$(d,p)$ ($c$), $^{34}$Cl$(d,p)$ $(d)$ and  $^{56}$Ni$(d,p)$ $(e,f)$ reactions and populating $l=0$ ($a,d$), $l=1$ ($f)$, $l=2$ ($b$) and $l=3$ ($c,e$) states calculated  at $E_d=$ 12 MeV  using the KD03 potential evaluated at a shifted by 57 MeV energy  
with (dashed) and without (solid) I3B terms in comparison with those obtained using standard Johnson-Tandy model with the same KD03 potential (dot-dashed).
}
\label{local_shift_comp_figure}
\end{figure*}

We found that the reduction in the real part of deuteron-target adiabatic  potentials that results from the inclusion of $\Delta E$ (roughly 15 MeV for the cases we investigate) produces a notable increase in the cross section. This is comparable in magnitude to the reduction caused by the addition of I3B terms. Some states display a greater sensitivity to the additions, with their impact most obvious for Ni calculations, but in general it was found that the reduction in cross section caused by the doubling of the imaginary potential offsets the effects of the energy shift.
This is demonstrated in Fig. \ref{local_shift_comp_figure} for a few typical cases. The factors that these cross sections differ by are denoted as $\sigma^{\text{I3B}}_{\text{KD03}}/\sigma_{\text{KD03}}$ and shown in Table \ref{states_table}. They can be as large as 25\%, while most fall within 5\% of cross sections generated using a standard approach, with little to no change found in angular distribution.

\begin{figure*}[htb]
\includegraphics[scale=0.24]{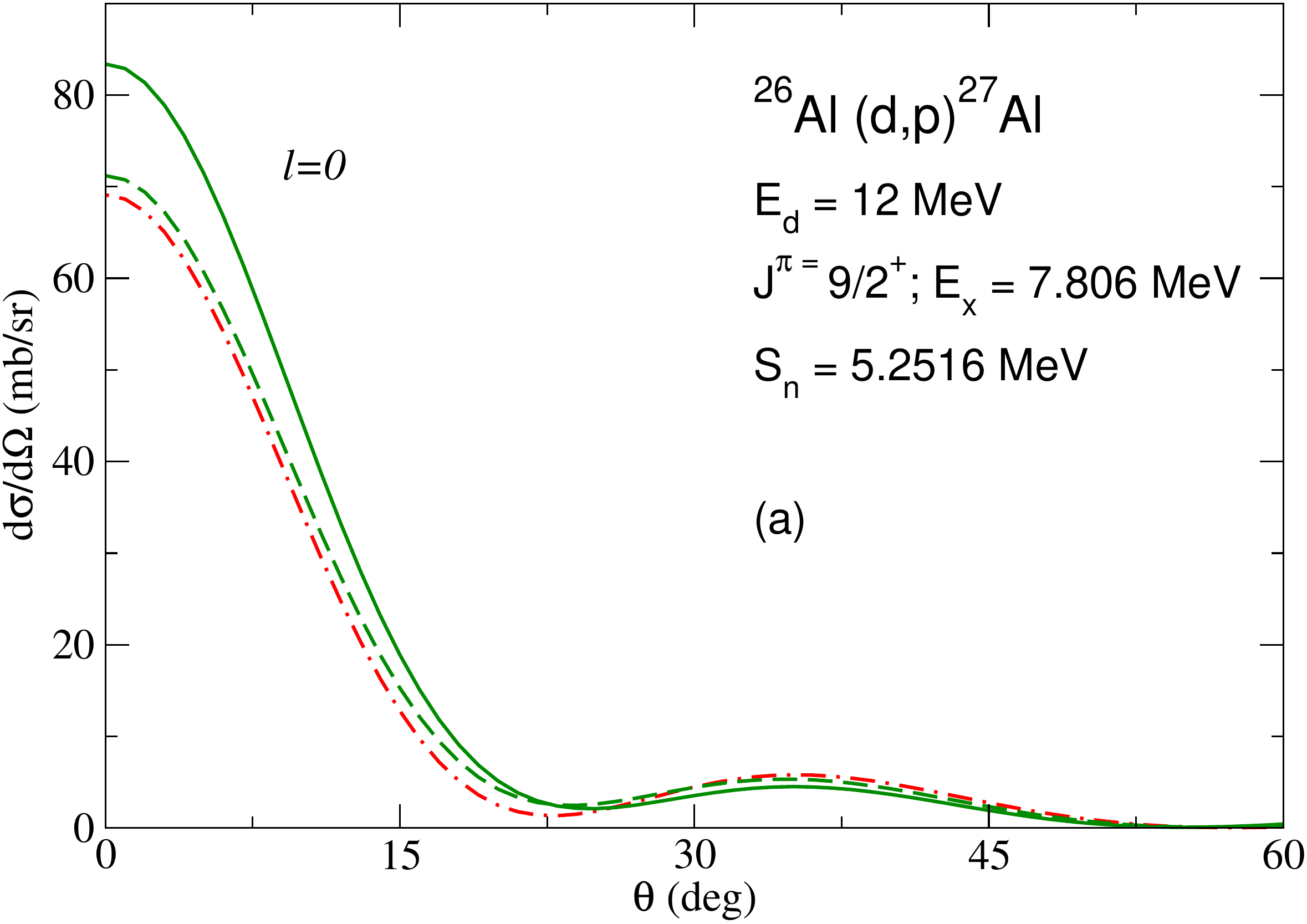}
\includegraphics[scale=0.24]{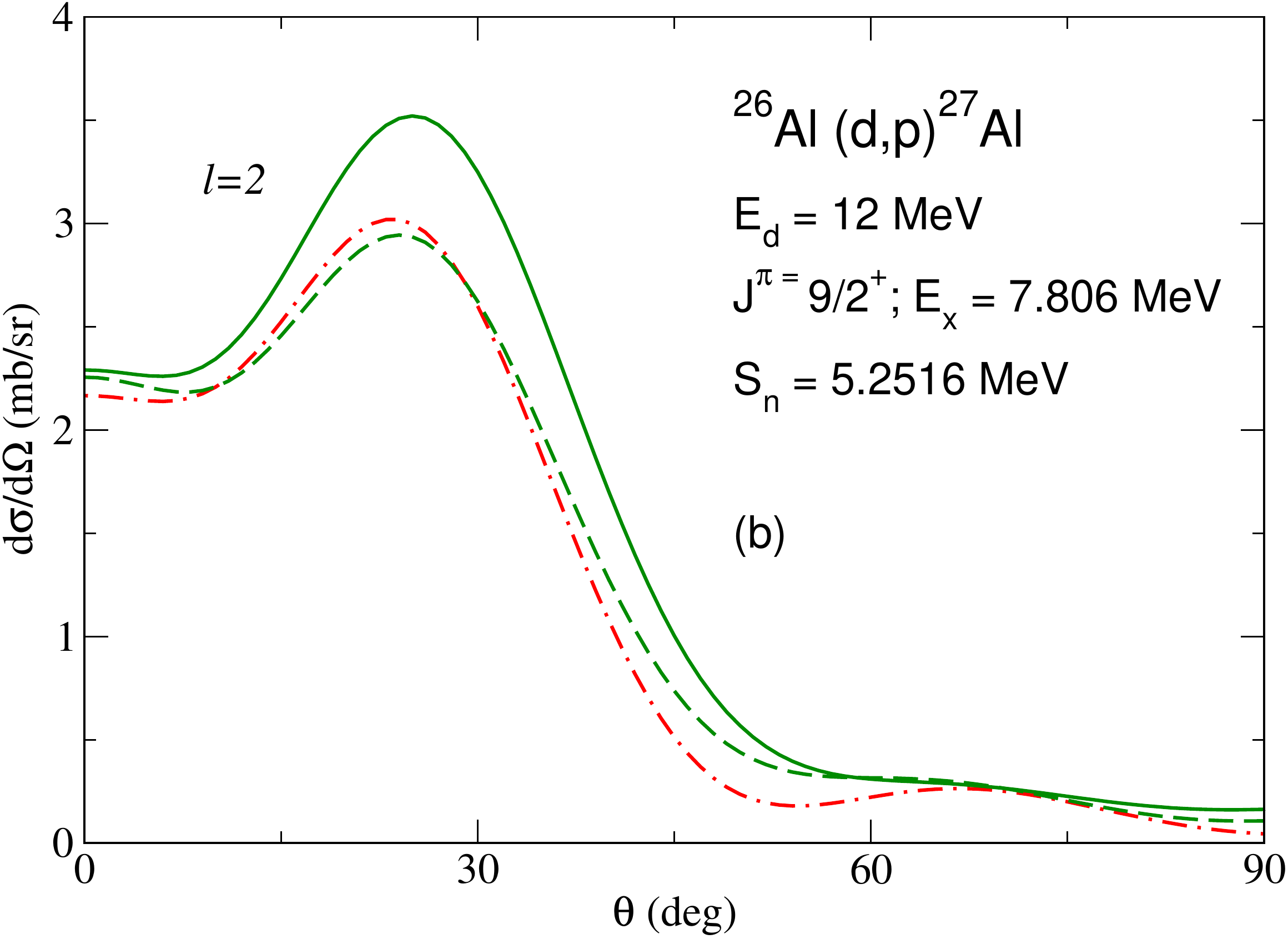}
\includegraphics[scale=0.24]{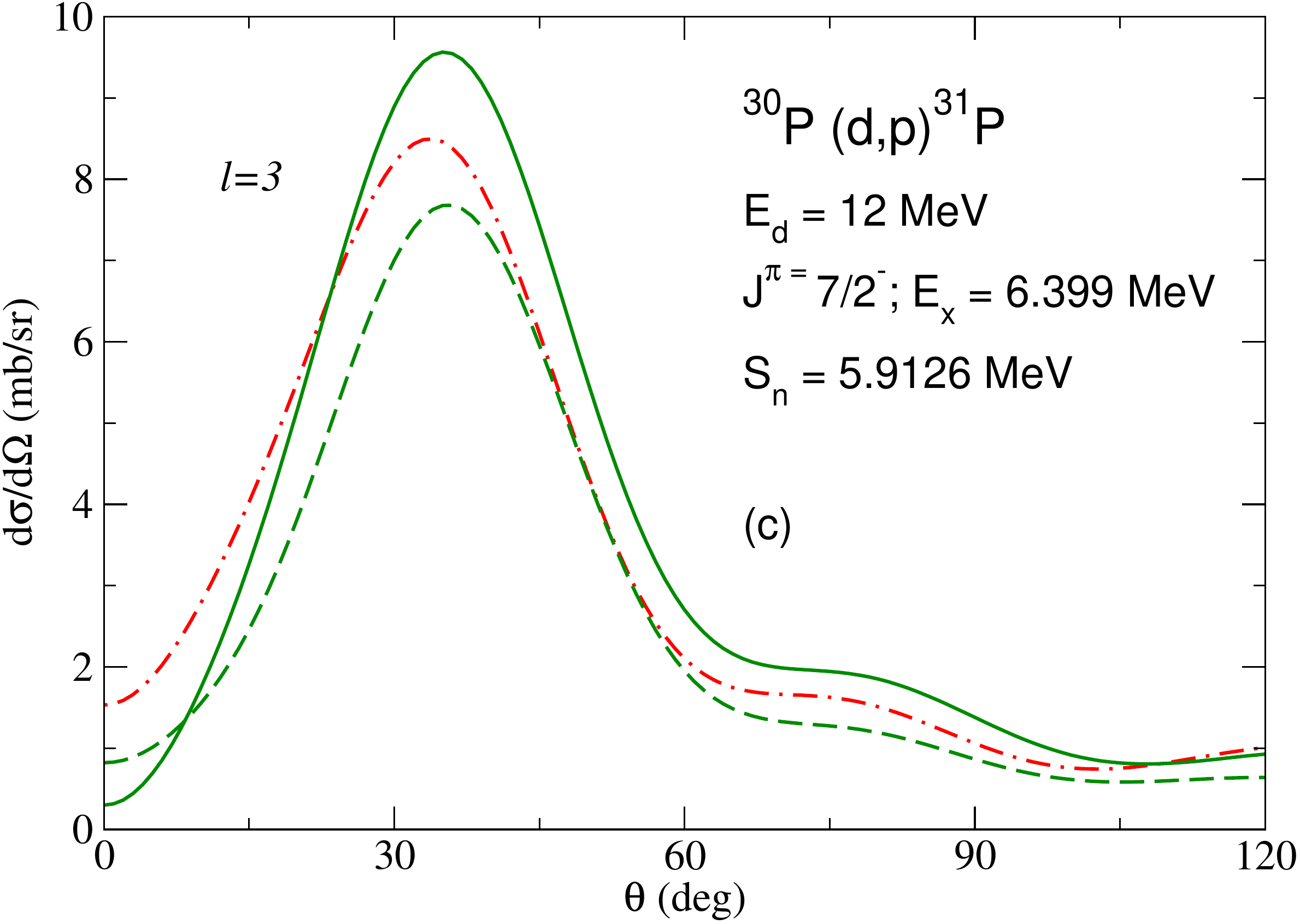}
\includegraphics[scale=0.24]{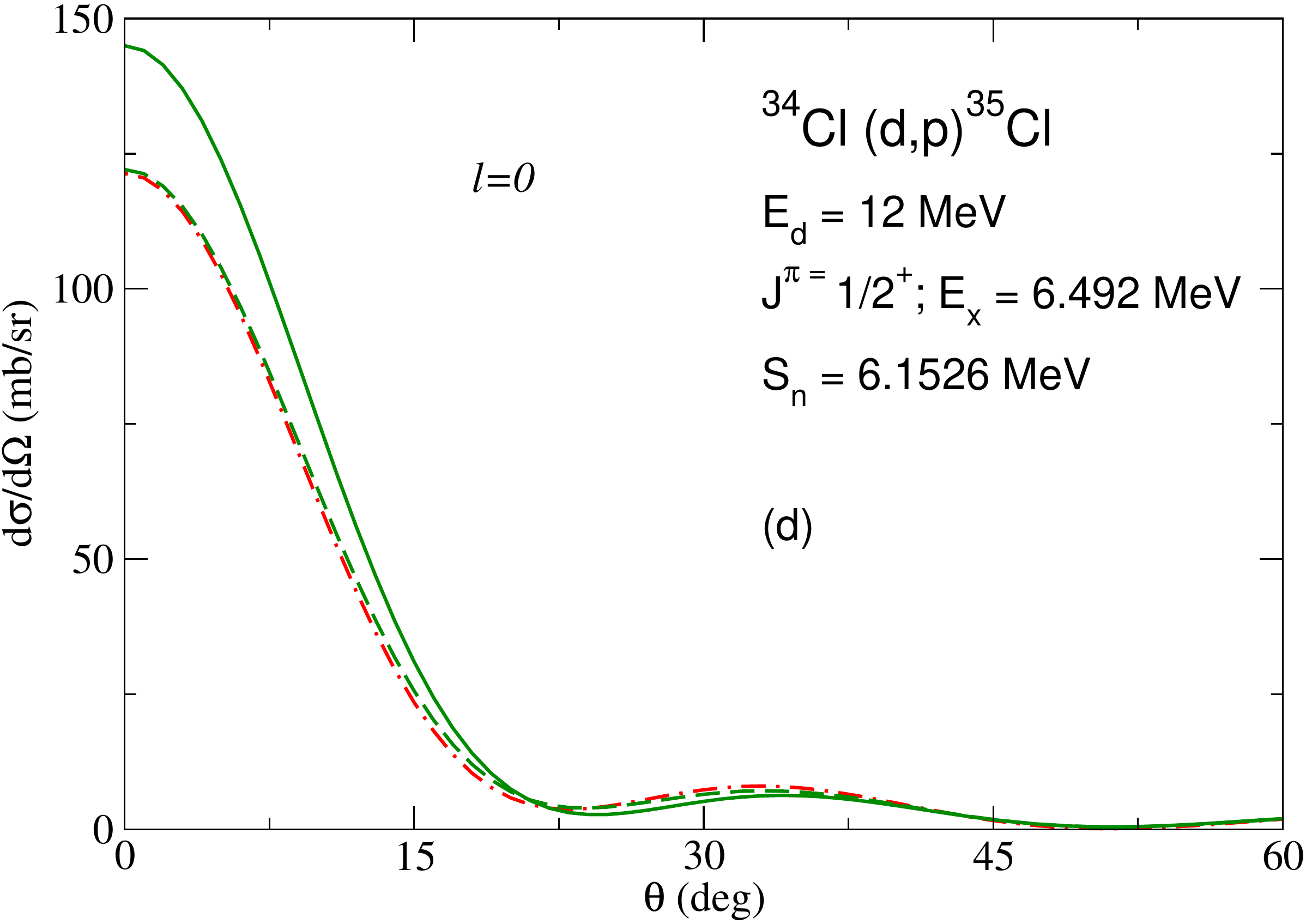}
\includegraphics[scale=0.24]{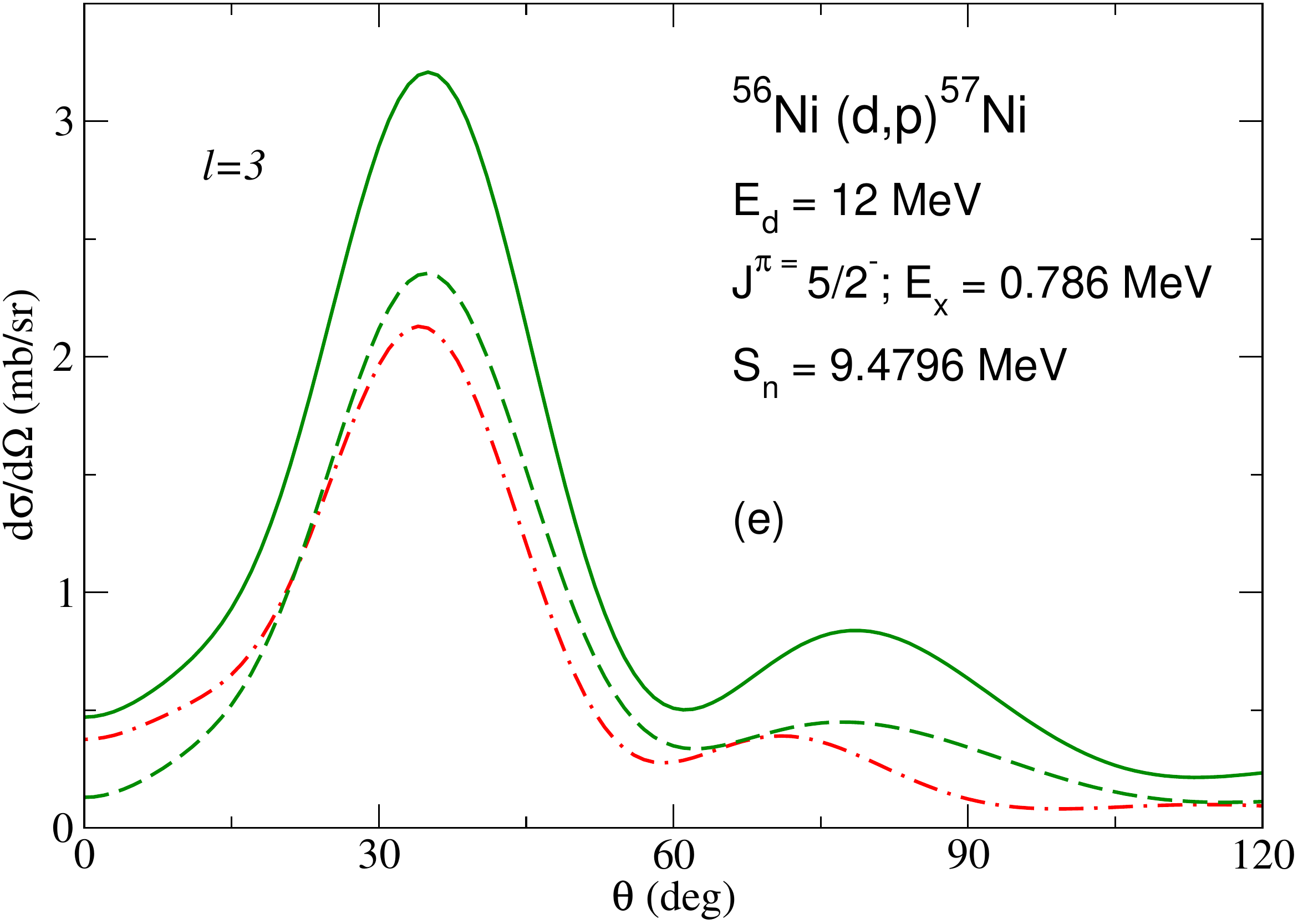}
\includegraphics[scale=0.24]{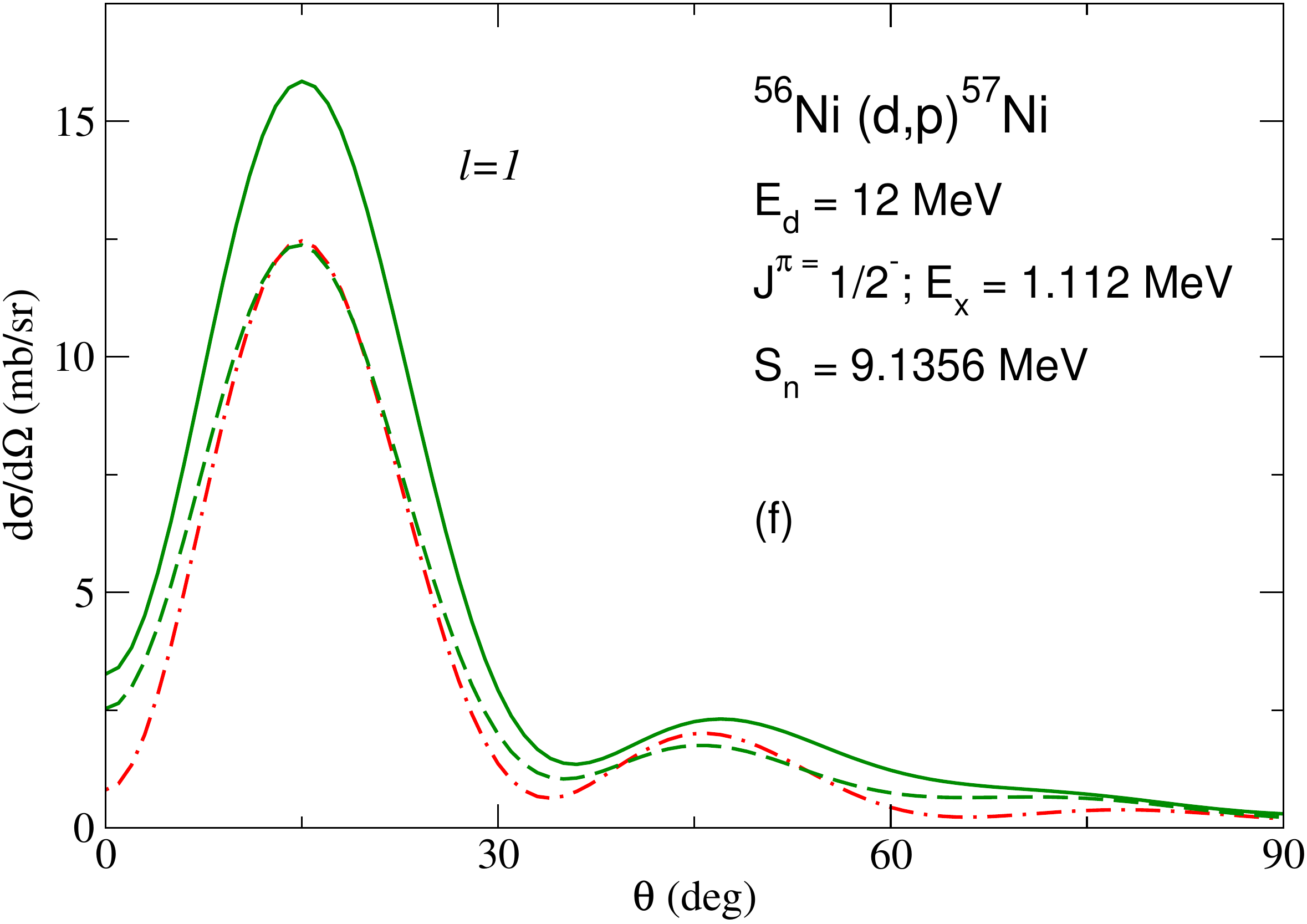}
\caption{
The ADWA  angular distributions for  $^{26}$Al$(d,p)$ $(a,b)$,  $^{30}$P$(d,p)$ ($c$), $^{34}$Cl$(d,p)$ $(d)$ and  $^{56}$Ni$(d,p)$ $(e,f)$ reactions and populating $l=0$ ($a,d$), $l=1$ ($f)$, $l=2$ ($b$) and $l=3$ ($c,e$) states calculated  at $E_d=$ 12 MeV  using the nonlocal equivalent of the KD03 potential  
with (dashed) and without (solid) I3B terms in comparison with those obtained using standard Johnson-Tandy model with the same KD03 potential (dot-dashed).}
\label{local_LEP_comp_figure}
\end{figure*}

It is known that local and nonlocal optical potentials give similar description of nucleon elastic scattering if the following transformation between the nonlocal potential and  its local equivalent $U_{\rm loc}^N$ is used \cite{PB},
\beq{}
U_{\rm loc}^N(r) = U_{NA}(r) \exp\left[ \frac{\mu \beta^2}{2\hbar^2}(U_{\rm loc}^N(r)-\bar{V}_c)\right],
\eeqn{Uloc}
where $\mu$ is the $N$-$A$ reduced mass and $\bar{V}_c$ is the Coulomb potential. This transformation assumes that the optical potentials are of the Perey-Buck form \cite{PB}
\beq
{\tilde U}_{NA}(E,r,r') = H(|\ve{r}-\ve{r}'|) U_{NA}(E,\frac{\ve{r}+\ve{r}'}{2}),
\eeqn{UNANL}
where $H(x)=\pi^{-3/2} \beta^{-3}\exp(-\frac{x^2}{\beta^2})$  and $\beta$  is some  non-locality range. A few recent examples of how this transformation performs and references to more literature on this subject can be found in \cite{Tim20a}.
We can assume that a known phenomenological energy-dependent optical potential $U_{\rm phen}(E,r)$ is a local-equivalent analog of an underlying nonlocal energy-dependent potential ${\tilde U}_{NA}(E,r,r')$, which we can restore through the  transformation (\ref{Uloc}).
Then according to Sec. II one should use  formfactors
$ U_{NA}(E,r)$ taken at the energy $E = E_{\rm eff} = E_d/2+ \Delta E$, treating  them as energy-independent. It was shown in \cite{Joh14} that this results in the local $d$-$A$ distorting potential, $U_{loc}$, being used in $(d,p)$ calculations, which is obtained from the transcendental equation 
\beq
U_{\rm loc}(E_d,r) \exp\left[- \frac{\mu_d \beta_d^2}{2 \hbar^2} U_{\rm loc}(E_d,r)\right] = 
{\cal V}(E_d,r),\,\,\,\,\,\,\,\,\,\,
\eeqn{Uloc2}
where 
\beq
{\cal V}(E_d,r) =  \exp\left[ \frac{\mu\beta^2}{2\hbar^2} E_{\rm eff}- \frac{ \mu_d\beta_d^2}{2\hbar^2}(E_d - {\bar V}_c)\right]
\eol \times M_0\left[
U_{\rm phen}^n( E_{\rm eff},r) \exp\left[- \frac{\mu \beta^2}{2\hbar^2}U_{\rm phen}^n(E_{\rm eff},r)\right] \right.
\eol \left. +
U_{\rm phen}^p(E_{\rm eff},r) \exp\left[ -\frac{\mu \beta^2}{2\hbar^2}(U_{\rm phen}^p(E_{\rm eff},r)-\bar{V}_c)\right] \right]. \eol
\eeqn{calV}
Here $\mu_d$ is the $d$-$A$ reduced mass,  $M_0$ is the value  that quantifies the overlap   of $\phi_1$, modified by nonlocality, with $\phi_d$ and $\beta_d$ is the deuteron effective nonlocality range. Exact definitions of $M_0$ and $\beta_d$ can be found in \cite{Tim13b}.

\newcolumntype{C}[1]{>{\centering\arraybackslash}p{#1}}
\begin{table*}[t]
\begin{center}
\caption{The ratio between total cross sections peaks calculated  from the first column at deuteron energies given in fifth column. The excitation energies (in keV), final state spins and quantum numbers of the populated level are given in the second, third and fourth column. In the following columns $\sigma_{\text{GRZ}}$ and  $\sigma^{\rm I3B}_{\text{GRZ}}$ denote  the cross sections calculated using GRZ  without and with I3B, respectively, $\sigma_{\rm KD03}$ are standard Johnson-Tandy results with KD03 potentials, $\sigma_{\rm KD03}^{\rm I3B}$ are obtained by applying shift $\Delta E$ and I3B effects to the local potential KD03 and $\sigma_{\rm KD03}^{\rm NLE,I3B}$ is obtained by restoring nonlocal equivalent of the local KD03 potential.}
\begin{tabular}{C{2.2cm}|C{1cm}C{1cm}C{1cm}|C{1.5cm}C{2cm}C{2cm}C{2.5cm}}
\hline
 & & & & & & & \\
Reaction & $E_x$   & $J^{\pi}$ & $lj$ & $E_d$ (MeV) & $\sigma^{\rm I3B}_{\text{GRZ}}/\sigma_{\rm KD03}$ & $\sigma^{\rm I3B}_{\text{KD03}}/\sigma_{\rm KD03}$ & $\sigma^{\rm NLE,I3B}_{\text{KD03}}/\sigma_{\rm KD03}$ \\ 
 & & & & & & & \\
\hline 
$^{26}$Al$(d,p)^{27}$Al & 3004 & 9/2$^{+}$ & $s_{\tfrac{1}{2}}$ & 12 & 0.886 & 1.187 & 1.169 \\
                        &      &           &                    & 60 & 0.725 & 1.112 & 1.239 \\
                        & 7806 & 9/2$^{+}$ & $s_{\tfrac{1}{2}}$ & 12 & 0.990 & 1.078 & 1.106 \\
                        &      &           &                    & 60 & 1.579 & 1.725 & 2.075 \\
                        & 7806 & 9/2$^{+}$ & $d_{\tfrac{3}{2}}$ & 12 & 1.027 & 1.180 & 1.187 \\
                        &      &           &                    & 60 & 0.557 & 1.013 & 1.073 \\
	                    & 7948 & 11/2$^{-}$& $p_{\tfrac{1}{2}}$ & 12 & 0.745 & 0.817 & 0.804 \\
	                    &      &           &                    & 60 & 0.458 & 0.623 & 0.640 \\
\hline
$^{30}$P$(d,p)^{31}$P   & 6336 & 1/2$^{+}$ & $s_{\tfrac{1}{2}}$ & 12 & 0.959 & 1.013 & 1.044 \\
                        &      &           &                    & 60 & 1.712 & 1.765 & 2.137 \\
                        & 6336 & 1/2$^{+}$ & $d_{\tfrac{3}{2}}$ & 12 & 1.068 & 1.155 & 1.173 \\
                        &      &           &                    & 60 & 0.654 & 1.101 & 1.178 \\
                        & 6399 & 7/2$^{-}$ & $f_{\tfrac{5}{2}}$ & 12 & 0.823 & 0.836 & 0.838 \\
                        &      &           &                    & 60 & 0.471 & 0.759 & 0.721 \\
\hline
$^{34}$Cl$(d,p)^{35}$Cl & 6492 & 1/2$^{+}$ & $s_{\tfrac{1}{2}}$ & 12 & 0.902 & 0.956 & 0.989 \\
                        &      &           &                    & 60 & 1.833 & 1.718 & 2.088 \\
                        & 6746 & 5/2$^{+}$ & $d_{\tfrac{5}{2}}$ & 12 & 0.857 & 0.867 & 0.906 \\
                        &      &           &                    & 60 & 0.921 & 1.184 & 1.265 \\
\hline
$^{56}$Ni$(d,p)^{57}$Ni & 768  & 5/2$^{-}$ & $f_{\tfrac{5}{2}}$ & 12 & 1.244 & 1.352 & 1.258 \\
                        &      &           &                    & 60 & 0.522 & 0.851 & 0.824 \\
                        & 1112 & 1/2$^{-}$ & $p_{\tfrac{1}{2}}$ & 12 & 1.126 & 1.231 & 1.165 \\
                        &      &           &                    & 60 & 1.050 & 1.089 & 1.188 \\
                        & 2443 & 5/2$^{-}$ & $f_{\tfrac{5}{2}}$ & 12 & 1.300 & 1.340 & 1.285 \\
                        &      &           &                    & 60 & 0.628 & 0.950 & 0.935 \\
                        & 2577 & 7/2$^{-}$ & $f_{\tfrac{7}{2}}$ & 12 & 1.065 & 1.029 & 1.026 \\
                        &      &           &                    & 60 & 0.935 & 1.167 & 1.164 \\
\hline
\end{tabular}   
\label{total_cs_table}
\end{center}{}
\end{table*}

We use the KD03 potential as $U_{\rm phen}^N$ and assume a standard value for the  nonlocality range $\beta=0.85$ fm, which in the Hulth\'en model gives  $M_0=0.78$ and $\beta_d = 0.4$ fm \cite{Tim13b}.  We calculate $U_{\rm loc}$ from Eqs.  (\ref{Uloc2})-(\ref{calV}) and use it in local $(d,p)$ calculations for the targets of previous subsections and compare the outcome with standard Johnson-Tandy KD03 calculations. A few typical angular distributions are shown in Fig. \ref{local_shift_comp_figure}. We have found that the angular distributions populating final $s$-, $p$- and $d$-wave states calculated with $U_{loc}$ are very similar to those obtained by directly applying the shift and I3B to the local KD03 potential with some differences seen for the cross sections to final $f$-wave state. In general, the differences increase with the orbital momenta of the transferred neutron. The ratios $\sigma^{\rm NLE,I3B}_{\rm KD03}/\sigma_{\rm KD03}$ of the cross sections in the maximum, obtained using these two different approaches employing local optical potentials systematics,  are shown in Table \ref{states_table}. Most results obtained with the two methods agrees within 3\%, the largest difference being 6\% for $^{57}$Ni(768 keV) state.

\section{Model uncertainties in total $(d,p)$ cross sections}

Usually, the determination of ANCs relies on a theoretical description of the angular distribution within the main maximum where the direct peripheral transfer mechanism is dominant. However, in some  inverse-kinematics-based experiments with radioactive beams this is not always possible. For example, detecting neutrons from $(d,n)$ reactions that could probe   $\Gamma_p$ directly can be very difficult. Also, deducing $(d,n)$ angular distributions by detecting the charged products of the resonance decay in coincidence could be very challenging due to the necessity of collecting a large proportion of resulting protons, having a good energy calibration and an absence of any contamination in the beam. 
For very narrow proton resonances of astrophysical interest lying close to the proton threshold, where $\gamma$-decay mode dominates, it has been possible to measure the total cross sections for a population of excited final states  by detecting   $\gamma$-rays emitted in their de-excitation  (see, for example, \cite{Al_dn,P_dn}). 
Total cross sections only were also measured in some $(d,p)$ experiments \cite{Ni_4,Hal21} without the corresponding differential cross sections.

The cross section at the main peak and the total cross section may carry different physical information. Indeed, the latter are obtained as
\beq
\sigma = \int_0^{2\pi} d\varphi \int_0^{\pi }  
\frac{d\sigma(\theta)}{d\Omega} \sin \theta d\theta ,
\eeqn{sigmatotal}
which contain $\sin\theta$  that suppresses the contribution from forward angles. 
This suppression could be particularly relevant for population of  $s$-wave final nucleon states, most interesting in astrophysical applications, especially for  higher incident energies. Increased contribution from a wider angular range, where other  mechanisms may come into play and where periphepherality may be lost, can lead to larger model uncertainties in the total cross sections than those established in the main peak.

In Table  \ref{total_cs_table} we show the various ratios of  total $(d,p)$ cross sections calculated in the same models that were used to obtain the results presented in Table \ref{states_table}. We show these ratios for two typical deuteron energies, $E_d = 12$ MeV, already studied in previous subsections and often used for measuring differential cross sections, and $E_d = 60$ MeV, which has been used several times to measure the total cross sections for $(d,n)$ reactions. 

For $E_d = 12$ MeV, the difference  between the total $\sigma^{\rm I3B}_{\rm GRZ}$ and $\sigma_{\rm KD03}$ cross sections is within 30\% . Unlike in the case of ratios of peak cross sections, where $\sigma^{\rm I3B}_{\rm GRZ}$ are either lower or similar to $\sigma_{\rm KD03}$,  the total cross section of $\sigma^{\rm I3B}_{\rm GRZ}$ could be either larger or smaller than those of $\sigma_{\rm KD03}$ ones. The ratios $\sigma^{\rm I3B}_{\rm GRZ}/\sigma_{\rm KD03}$ of total cross sections are larger than the corresponding ratios of peak cross sections, thus confirming that the model uncertainties in the total cross sections are generally higher than those in the peak values. 
Similar situation occurs when local potential KD03, modified to account for three-body effects, is used in calculations. Different treatment of this modification, either applied directly  or using local potential reconstruction techique of Eqs. (\ref{Uloc2})-(\ref{calV}) via nonlocal equivalent of KD03, gives the total cross sections differing by no more  6\%, which is marginally larger than the difference of 2\% between that obtained by the two methods for peak cross sections.

\begin{table*}[t]
\begin{center}
\caption{Squared ANCs $C_n^2$ (in fm$^{-1}$)  and widths $\Gamma_p$  (in eV) for mirror pairs shown in first column. The second and third column displays resonance energies $E_R$ (in MeV) and nucleon orbital momentum $l$. The calculations are shown for standard ADWA with KD03 (4th and 5th column) and for including the I3B effects in nonlocal, GRZ, and local, KD03,  optical potentials. The I3B effect in the latter was introduced either directly (8th and 9th column) or in its nonlocal equivalent (10th and 11th columns). }
\begin{tabular}{C{3.5cm}C{0.6cm} C{0.6cm} |C{1.0cm}C{1.6cm}C{1.0cm}C{1.6cm}C{1.0cm}C{1.6cm}C{1.0cm}C{1.5cm}}
\hline
 & & & \multicolumn{2}{c}{KD03} & \multicolumn{2}{c}{GRZ+I3B} & \multicolumn{2}{c}{KD03(L)+I3B} & \multicolumn{2}{c}{KD03(NLE)+I3B} \\
 Mirror pair &  $E_R$ & $l$ & $C^2_n$ & $\Gamma_p$& $C^2_n$ & $\Gamma_p$& $C^2_n$ & $\Gamma_p$& $C^2_n$ & $\Gamma_p$ \\ 
\hline
$^{27}$Al(7806) - $^{27}$Si(7590) & 127 & 0  & 0.258  & 6.27$\cdot$10$^{-8}$ & 0.365 &8.88$\cdot$10$^{-8}$ & 0.247& 6.01$\cdot$10$^{-8}$ & 0.250 & 6.08$\cdot$10$^{-8}$ \\
$^{27}$Al(7806) - $^{27}$Si(7590) & 127 & 2  & 0.148  & 2.15$\cdot$10$^{-9}$ & 0.128 &1.86$\cdot$10$^{-9}$ & 0.125& 1.81$\cdot$10$^{-9}$ & 0.120 & 1.74$\cdot$10$^{-9}$ \\
$^{27}$Al(7948) - $^{27}$Si(7652) & 189 & 1  & 0.650  & 3.63$\cdot$10$^{-5}$ & 0.979 &5.47$\cdot$10$^{-5}$ & 0.793& 4.43$\cdot$10$^{-5}$ & 0.797 & 4.45$\cdot$10$^{-5}$ \\
$^{57}$Ni(768) - $^{57}$Cu(1028) & 336 & 3  & 23.2  & 1.24$\cdot$10$^{-11}$ & 15.2 & 8.14$\cdot$10$^{-12}$ & 18.9& 1.01$\cdot$10$^{-11}$ & 19.8 & 1.06$\cdot$10$^{-11}$ \\
$^{57}$Ni(1112) - $^{57}$Cu(1106) & 416 & 1  & 202  & 2.64$\cdot$10$^{-7}$ & 163 &2.13$\cdot$10$^{-7}$ & 203& 2.65$\cdot$10$^{-7}$ & 203 & 2.65$\cdot$10$^{-7}$ \\
\hline
\end{tabular} \label{Gammap}  
\end{center}{}
\end{table*}

The calculations at $E_d = 60$ MeV revealed that the cross sections are much more sensitive to the model choice for $V_{np}$, to extent that this choice can strongly affect the shapes of the angular distributions (not shown here). Similar effect has been observed in \cite{Gom18} for $^{26m}$Al$(d,p)^{27}$Al at $E_d = 50$ MeV. As in previous subsections, we use the Hulth\'en model, which in the case of nonlocal potentials suppresses the contribution from poorly-defined high $n$-$p$ momenta causing strong deviation of the ADWA cross sections   from  exact  three-body dynamics predictions. Unlike in the case of $E_d = 12$ MeV, applying the energy shift and doubling imaginary part can dramatically affect the shapes of  angular distributions so that they may lose resemblance with the standard ADWA predictions.  Significant shape changes make comparison of peak cross sections meaningless. We calculate the ratio of the total cross sections obtained with $\Delta E$ and I3B force to  the standard ADWA ones showing   them in Table  \ref{total_cs_table}. As in the case of $E_d = 12$ MeV, we apply the shift $\Delta E$ and doubling imaginary part to nonlocal potential  GRZ and to local potential KD03, both directly or to its nonlocal equivalent. The three-body effects introduced in this way have a stronger influence on the total cross sections than in the 12 MeV case, in particularly for populating $s_{1/2}$ states, as expected. This influence is different depending of what kind of potential was used in the calculations, nonlocal GRZ or local KD03 corrected for I3B effects. In general, the model uncertainty due to the three-body nature of optical potentials can be as large as a factor of two. It can result both in an increase and a decrease of the cross sections depending on the reaction choice.



\section{Neutron ANCs and proton widths of mirror states }

To illustrate how model uncertainties due to three-body nature of optical potentials in the $A+n+p$ system affect information about reactions in stellar environments we have determined ANCs for $^{27}$Al$^*(7806,7948)$ and $^{57}$Ni$^*(768,1112)$ states from the $(d,p)$ angular distributions measured at $E_d = 12$ MeV and 8.9 MeV in \cite{Al_dp2} and \cite{Ni_3}, respectively, and then calculated the widths of mirror proton resonances using relation 
 \cite{Tim03a}
\begin{equation}
\dfrac{\Gamma_p}{C_n^2}\approx {\cal R}_{res} \equiv  \dfrac{\kappa_p}{\mu}\left|\dfrac{F_l(\kappa_p R_N)}{\kappa_p R_Nj_l(i\kappa_n R_N)}\right|^2,
\label{ANC_Gamma_link}
\end{equation}
where $C_n$ is the neutron ANC, $\kappa_{(n,p)}=\sqrt{2\mu \varepsilon_{(n,p)}/\hbar^2}$, $\mu$ is the $A$-$N$ reduced mass, $\varepsilon_{(n,p)}$ is the (positive) separation for neutron or proton resonance energy, $R_N= R_0 A^{1/3}$ is the range of internal region  of $A$, $j_l$ is a spherical Bessel function and $F_l$ is the regular Coulomb wave function. 
We have chosen $R_0=1.3$ fm and checked that for all our cases the ratio ${\cal R}_{res}$  changes by less than 2\% for $1.25 \leq R_0 \leq 1.35 $ fm. We should note that Eq. (\ref{ANC_Gamma_link}) is a model-independent approximation suitable for fast calculations as it requires knowledge of nucleon energies and target charges only. Since we investigate dependence of neutron ANCs and mirror proton widths on optical potentials only, for the purpose of our paper, the exact value of $\Gamma_p/C^2_n$ is not needed. However, it can be obtained in terms of the Wronskians from the radial overlap functions and regular solutions of the two-body Schrödinger equation with the short-range interaction excluded (see \cite{Muk19} for details). 

We first checked the extent to which these reactions are peripheral by changing the radius, $r_0$, of the bound neutron potential within the range of 1.1-1.4 fm and comparing the calculated cross sections to the experimental ones. In all cases the spectroscopic factors changed much more strongly than the ANCs with varying $r_0$, which is a good indication that the reactions considered are mainly sensitive to the peripheral region of the nucleus. The strongest spread in ANC squared, up to 14\%, was observed for populating $^{57}$Ni$^*(1112)$. However, in this case the spectroscopic factor changed by 43\% with respect to the value obtained with standard radius $r_0 = 1.25$ fm.
Since our main aim is to quantify uncertainties due to the I3B effects, we ignore the uncertainties in ANCs due to residual dependence on $r_0$ as well as  experimental uncertainties. 
We   note that for the $^{27}$Al(7806) state the cross section is a sum of two terms corresponding to $l=0$ and $l=2$ transfers. This creates additional uncertainties for ANCs determined for this particular state.

The ANCs squared, proton widths of mirror states for GRZ and two versions of corrected KD03 potential are shown in Table \ref{Gammap}. Results obtained with standard ADWA employing the same KD03 are shown there as well. One can see that introducing I3B effects  gives different results depending on nuclear state and on what kind of potential is employed, local or nonlocal. The I3B effect can work both ways, either increasing of decreasing 
$C_n^2$ and the corresponding  $\Gamma_p$ values, in the worst case - $^{57}$Ni(768) state with GRZ - deviating from the standard ADWA approach by 52\%. Accounting for I3B force using local optical potential KD03 in most cases does not exceed  22\% change with respect to ADWA, apart from the $l=2$ case in $^{26}$Al(7806) where uncertainty of extracted $C_n^2$ is expected to be large. In all cases two different ways of introducing I3B effects based on KD03 local potential  give  very similar results. Introducing I3B force into nonlocal potential GRZ can cause stronger effect, up to 52\% in $^{57}$Ni(1112).

The widths $\Gamma_p$ shown in Table \ref{Gammap} have already been determined using mirror symmetry assumptions in previous works from $(d,p)$ angular distributions \cite{Al_dp2,Ni_3}. But these widths were estimated using a different model which assumes that $\Gamma_p$ is given by product of Coulomb barrier penetrability factor, doubled  Wigner single-particle width and proton spectroscopic factor, equal to that of the  mirror neutron. While the assumption about equality of  spectroscopic factors in mirror states should be good, the Coulomb barrier penetrability depends on the geometry of the nuclear potential, which  introduces additional, usually not considered, uncertainties in determination $\Gamma_p$ from mirror symmetry. Therefore, we can expect differences in $\Gamma_p$ determined by traditional Coulomb penetrability method and the method of Eq. (\ref{ANC_Gamma_link}) used here.  We believe that a systematic comparison between the two method should be a subject of a separate paper. We just note that the width $\Gamma_p(l=0) = 2.5\cdot10^{-8}$ eV  of the $E_R = 127$ keV  resonance in $^{27}$Si obtained in \cite{Al_dp2} is about 3 times  smaller than the values obtained in this paper while the 
$\Gamma_p = 5.1 \cdot10^{-5}$ eV width for  $E_R = 188$   keV resonance in $^{27}$Si    is similar to our results. For $^{57}$Cu our $\Gamma_p$ widths are about 100\% and 40\% larger than the $\Gamma_p = 5.7 \cdot10^{-12}$ eV and 1.9$ \cdot10^{-7}$ eV values obtained in \cite{Ni_4} 
for $E_R = 336$ and 416 KeV resonances, respectively.

Finally, we have to comment on the determination of $\Gamma_p$ from total $(d,p)$ cross sections $\sigma_t$. We find that at $E_d = 12$ MeV the calculated total cross sections are not very sensitive to changes in $r_0$, which suggests that peripherality of these reactions is lost, but offers an opportunity of determining the spectroscopic factors. The ANCs $C_n^2$ obtained from total cross sections could have large uncertainties, between 30 to 100\%, being either comparable to or even larger than those due to the I3B effects. At $E_d = 60$ MeV the situation is different. Apart from $^{27}$Al(7806) with $l=0$ and $^{57}$Ni(2577) cases,
the ratio $\sigma_t/b^2$, where $b$ is the single-particle ANC,  changes significantly with $r_0$, indicating that peripherality is also lost at higher energies, as expected. At the same time $\sigma_t$ also becomes $r_0$-dependent, inducing large uncertainties in the spectroscopic factors as well. Once again, these uncertainties can be comparable to those originating from introducing I3B effects. A special case is $^{27}$Al(7806) with $l=0$ and $^{57}$Ni(2577) cases, where changing $r_0$ within the interval of the most probable values, between 1.15 and 1.35 fm, results in a 3\%  change in $\sigma_t/b^2$, with respect to the average value. This is sufficient to deduce $C_n^2$ for $^{57}$Ni(2577), where only one $l$-value contributes to the total cross section. In this case, using $\sigma_t = 1.24$ mb for $^{56}$Ni$(d,p)^{57}$Ni(2577) reaction measured at $E_d = 64$ MeV we obtain $\Gamma_p = 2.68$ eV, which is about 5 times larger than the 0.53 eV value obtained via Coulomb barrier penetrability method, with uncertainties due to I3B effects being  6-16\%. 
For $^{27}$Al(7806) the  ANCs and the corresponding $\Gamma_p$ cannot be determined since two unknown values need to be determined from only one experimental observable.


\section{Conclusions}

We have presented calculations for  cross sections of several $(d,p)$ reactions populating excited states in $^{27}$Al, $^{31}$P, $^{35}$Cl and $^{57}$Ni, which are mirror analogs of important resonances contributing to nucleosynthesis in various stellar environments via $rp$-process. The main aim of these calculations was to quantify model uncertainties of the widely used ADWA arising due to projecting target  excitations out, which affects two-body nucleon optical potentials in  the $A+n+p$ three-body system and induces multiple scattering terms in the three-body optical potential. Within ADWA,   these effects are accounted for by  $(a)$ evaluating the $n$-$A$ and $p$-$A$ optical potentials at   energy  shifted with respect to the $E_d/2$ value by half the $n$-$p$ kinetic energy within the short range of $V_{np}$, and $(b)$ doubling their imaginary parts.  These corrections are introduced into two different nucleon optical potentials, the nonlocal energy-dependent potential GRZ and local global potential KD03, treating the latter in two different ways. The calculations were carried out at two typical energies,  available at radioactive beam facilities, that have been already used for indirect study of astrophysically relevant reactions.

The shift of energy at which optical potentials should be evaluated in ADWA   strongly depends on the choice of the deuteron model and this can significantly affect the $(d,p)$ cross sections. 
The strong sensitivity to the deuteron model arises due to the  existence of a deuteron $d$-wave state  that contributes about 40\%  to the  $n$-$p$ potential energy matrix element.  Introducing I3B effects by doubling the imaginary part of the optical potentials reduce dependence on deuteron model choice. Choosing the $s$-wave Hulth\'en model we reduce the contribution from  high $n$-$p$ momenta along with contributions from non-adiabatic effects.

With the chosen deuteron model, 
including the three-body nature of the optical potentials in the $A+n+p$ system 
at a typical deuteron energy of $E_d = 12$ MeV does not change the shape of the angular distributions significantly with respect to predictions of the standard ADWA, resulting in a renormalization of the cross sections. This renormalization depends on the choice of optical potential being distinct for nonlocal and local ones. For nonlocal potentials the I3B effects reduce the cross sections by up to 40\% for $^{26}$Al, $^{30}$P and $^{34}$Cl targets, while for $^{56}$Ni this reduction does not exceed 20\%. For local potentials the change likewise does not exceed 20\%, but can go either way, increasing or decreasing the cross sections. Two different ways of applying I3B effects in local potentials, either directly or to restored nonlocal equivalent, give very similar results.
The total cross sections have larger model uncertainties, especially at higher deuteron energies. While at $E_d = 12$ MeV they do not exceed 35\%, being in most cases within 10-20\%.  At $E_d = 60$ MeV this uncertainty can be as high as 100\% causing changes both ways, increasing or decreasing. These uncertainties will directly propagate to corresponding uncertainties in the widths of astrophysically relevant resonances and the associated proton capture reaction rates.

Current estimates of three-body effects in the optical potential of the $A+n+p$ system, arising from projecting out target excitations, are based on ADWA. It is important to understand how they may change
beyond the adiabatic assumptions.   The only attempt to do this has been carried out using Faddeev equations with excited states, but neglecting the many-body nature of the target. Similar to our findings, treating the energy-dependence of the optical potentials explicitly by adjusting two-body Faddeev $T$-matrices increases  the $(d,p)$ cross sections requiring three-body optical potential that provides increasing absorption \cite{Del17}. This is qualitatively the same effect as the one considered here. It is important to extend the three-body optical potential study to many-body systems beyond the adiabatic approximation. This will help to build more accurate models that will help to extract reliable spectroscopic information  from $(d,p)$ reactions for various applications, including understanding of stellar nucleosynthesis.

Finally, we would like to mention that in addition to three-body potential effects induced by projecting out target excitations, there should exist a contribution to the three-body Hamiltonian that arises due to three-nucleon  interactions between $n$, $p$ and individual nucleons in the target \cite{Tim20}. This contribution depends on the  choice of target, deuteron incident energy, binding energy of the state populated by the transferred neutron and its orbital momentum. The three-nucleon  interaction also introduces additional term into $(d,p)$ T-matrix \cite{Tim18}. These effects can introduce further model uncertainty to $(d,p)$ cross sections, which will affect the spectroscopic information obtained from them. More work is needed to clarify their role.

\section*{Acknowledgements.}
 This work was supported by the United Kingdom Science and Technology Facilities Council (STFC) under Grant No. ST/L005743/1.


\begin{thebibliography}{1}
\bibitem{Tim03a}  N. K. Timofeyuk, R. C. Johnson and A. M. Mukhamedzhanov, Phys. Rev. Lett. {\bf 91}, 232501 (2003).
\bibitem{JT} R.C. Johnson and P.C. Tandy, Nucl. Phys. A \textbf{235}, 56 (1974).
\bibitem{Joh14} R.C. Johnson and N.K. Timofeyuk, Phys. Rev. C \textbf{89}, 024605 (2014).
\bibitem{Din19} M.J. Dinmore, N.K. Timofeyuk, J.S. Al-Khalili and R.C. Johnson, Phys. Rev. C \textbf{99}, 064612 (2019)
\bibitem{Tim20} N.K. Timofeyuk, M.J. Dinmore, and J.S. Al-Khalili, Phys. Rev. C \textbf{102}, 064616 (2020).
\bibitem{Bai16}  G.W. Bailey, N.K. Timofeyuk, J. A. Tostevin, Phys. Rev. Lett. \textbf{117}, 162502 (2016).
\bibitem{Wal16} S.J. Waldecker and N.K. Timofeyuk, Phys. Rev. C \textbf{94}, 034609 (2016).


\bibitem{Al_1}G. Lotay, P.J. Woods, D. Seweryniak, M. P. Carpenter, R. V. F. Janssens, and S. Zhu, Phys. Rev. C \textbf{80}, 055802 (2009).
\bibitem{Al_dp1} G.Lotay, M.Moukaddam, et al., Eur. Phys. J. A \textbf{56}, 3 (2020).
\bibitem{Al_dp2} V. Margerin, G. Lotay, P.J. Woods, et al., Phys. Rev. Lett. \textbf{115}, 062701 (2015).
\bibitem{Al_dn} A. Kankainen, P. Woods, F. Nunes, et al., Eur. Phys. J. A \textbf{52}, 1 (2016).
\bibitem{P_1} J. Fallis, A. Parikh, P.F. Bertone, S. Bishop, L. Buchmann,A.A. Chen
et al., Phys. Rev. C, \textbf{88}, 045801 (2013).
\bibitem{iliadis2002} C, Iliadis et al., Astrophys. J. Suppl. Ser. \textbf{142}, 105 (2002).
\bibitem{P_dn} A. Kankainen et al., Physics Letters B \textbf{769}, 549 (2017)
\bibitem{Cl_1} J. Jose, et al., Astrophys. J. \textbf{612}, 414 (2004)
\bibitem{Ni_1} A.E. Champagne and M. Wiescher, Annu. Rev. Nucl. Part. Sci. \textbf{42}, 39 (1992)
\bibitem{Ni_2} J.W. Truran, et al., J. Phys.: Conf. Ser. \textbf{337}, 012040 (2012)
\bibitem{Ni_3} K.E. Rehm, F. Borasi,C.L. Jiang,D. Ackermann,I. Ahmad, B.A. Brown, F. Brumwell 
et al., Phys. Rev. Lett., \textbf{80}, 676 (1998)
\bibitem{Ni_4} D. Kahl, et al., Physics Letters B \textbf{797}, 134803 (2019)


\bibitem{twofnr} J.A. Tostevin, University of Surrey version of the code {\sc twofnr} (of M. Toyama, M. Igarashi and N. Kishida) and code {\sc front} (private communication).
\bibitem{Bai17} G.W. Bailey, N.K. Timofeyuk, J.A. Tostevin, Phys. Rev. C \textbf{95}, 024603 (2017).
\bibitem{GRZ} M.M. Giannini, G. Ricco, A. Zucchiatti, Ann. Phys. \textbf{124}, 208 (1980).

\bibitem{Hulthen} L. Hulth\'en and M. Sugawara, The Two-Nucleon Problem (Springer-Verlag, Berlin, 1957)
\bibitem{AV18} R. B. Wiringa, V. G. J. Stoks, and R. Schiavilla, Phys. Rev. C \textbf{51}, 38 (1995)
\bibitem{CDBonn} R. Machleidt, Phys. Rev. C \textbf{63}, 024001 (2001)
\bibitem{N4LO} E. Epelbaum, H. Krebs, and U.-G. Mei\ss ner, Phys. Rev. Lett. \textbf{115}, 122301 (2015)

\bibitem{Fry15} C. Fry et al Phys. Rev. C 91, 015803(2015)
\bibitem{mahauxsartor} C. Mahaux, R. Sartor, Nucl. Phys. A, \textbf{484}, 205 (1988).
\bibitem{Del18} A. Deltuva,  Phys.Rev. C \textbf{98}, 021603(R) (2018).
\bibitem{Gom18} M. G\'omez-Ramos and N.K. Timofeyuk, Phys. Rev. C \textbf{98}, 011601(R) (2018).
\bibitem{Tim13b} N.K. Timofeyuk and R.C. Johnson, Phys. Rev. C87, 064610 (2013)
\bibitem{PB} F. Perey and B. Buck, Nucl. Phys. 32, 353 (1962).
\bibitem{Tim20a} N.K. Timofeyuk and R.C. Johnson, Prog. Part. Nucl. Phys. {\bf 111}, 103738 (2020)
\bibitem{KD03} A.J. Koning and J.P. Delaroche, Nucl. Phys. A, \textbf{713}, 231 (2008)
\bibitem{Hal21} S. Hallam, G. Lotay, A. Gade, D.T. Doherty, J. Belarge, P.C.Bender 
et al,   Phys. Rev. Lett. 126, 042701 (2021)
\bibitem{Muk19} A. M. Mukhamedzhanov, Phys. Rev. C 99, 024311  (2019)
\bibitem{Del17} A. Deltuva, D. Jurciukonis, and E. Norvaisas, Phys. Lett. B 769, 202 (2017)
\bibitem{Tim18} N.K. Timofeyuk, Phys. Rev. C {\bf 97}, {054601},  (2018)

\end{thebibliography}
\end{document}